\documentclass[onecolumn, draftcls, 10pt]{IEEEtran}

\usepackage{amsmath,amssymb,diagbox}
\usepackage{subfig,url, cite}
\usepackage{epsfig}
\usepackage{epstopdf}
\usepackage{float, cite}
\usepackage{amsfonts,amsthm}		
\usepackage{color}
\usepackage{caption}
\usepackage{algorithm}
\usepackage{algorithmic, bm}

\newcommand{\bmath}[1]{\mathbf{\bm{#1}}}
\def \bx{\bmath{x}}
\def \bA{\bmath{A}}
\def \ba{\bmath{a}}
\def \by{\bmath{y}}

\theoremstyle{plain}
\newtheorem{thmi}{Theorem}[section]

\newtheorem{cori}{Corollary}[section]

\newtheorem{lemmai}{Lemma}[section]

\title{\huge On the Fundamental Limits of Recovering Tree Sparse Vectors from Noisy Linear Measurements}

\author{Akshay Soni, ~\IEEEmembership{Student Member, ~IEEE} and Jarvis Haupt, ~\IEEEmembership{Member, ~IEEE}\thanks{Manuscript submitted June 18, 2013; revised October 15, 2013. The authors are with the Department of Electrical and Computer Engineering, University of Minnesota -- Twin Cities, Minneapolis, MN, 55455 USA; emails: {\tt \{sonix022, jdhaupt\}@umn.edu}. A portion of this work appeared at the 2011 IEEE Asilomar Conference on Signals, Systems, and Computers, and a shorter summary version of this paper appeared at the 2013 IEEE Global Conference on Signal and Information Processing. This work was supported by DARPA/ONR Award No. N66001-11-1-4090. 

Copyright $\copyright$ 2013 IEEE. Personal use of this material is permitted. However, permission to use this material for any other purposes must be obtained from the IEEE by sending request to \tt{pubs-permissions@ieee.org}.}}

\begin{document}

\maketitle

\begin{abstract}
Recent breakthrough results in compressive sensing (CS) have established that many high dimensional signals can be accurately recovered from a relatively small number of non-adaptive linear observations, provided that the signals possess a sparse representation in some basis.  Subsequent efforts have shown that the performance of CS can be improved by exploiting additional structure in the locations of the nonzero signal coefficients during inference, or by utilizing some form of data-dependent adaptive measurement focusing during the sensing process.   To our knowledge, our own previous work was the first to establish the potential benefits that can be achieved when fusing the notions of adaptive sensing and structured sparsity -- that work examined the task of support recovery from noisy linear measurements, and established that an adaptive sensing strategy specifically tailored to signals that are tree-sparse can significantly outperform adaptive and non-adaptive sensing strategies that are agnostic to the underlying structure. In this work we establish fundamental performance limits for the task of support recovery of tree-sparse signals from noisy measurements, in settings where measurements may be obtained either non-adaptively (using a randomized Gaussian measurement strategy motivated by initial CS investigations) or by any adaptive sensing strategy.  Our main results here imply that the adaptive tree sensing procedure analyzed in our previous work is nearly optimal, in the sense that no other sensing and estimation strategy can perform fundamentally better for identifying the support of tree-sparse signals.  
\end{abstract}

\begin{IEEEkeywords}
Adaptive sensing, compressive sensing, minimax lower bounds, sparse support recovery, structured sparsity, tree sparsity. 
\end{IEEEkeywords}

\section{Introduction}\label{sec:intro}

In recent years, the development and analysis of new sampling and inference methods that make efficient use of measurement resources has received a renewed and concentrated focus. Many of the compelling new investigations in this area share a unifying theme -- they leverage the phenomenon of \emph{sparsity} as a means for describing inherently simple (i.e., low-dimensional) structure that is often present in many signals of interest.   

Consider the task of inferring a (perhaps very high-dimensional) vector $\bx\in\mathbb{R}^n$.  Compressive sensing (CS) prescribes collecting non-adaptive linear measurements of $\bx$ by ``projecting'' it onto a collection of $n$-dimensional ``measurement vectors.''  Formally, CS observations may be modeled as
\begin{equation}\label{eqn:obs}
y_j = \langle \ba_j, \bx \rangle + w_j = \ba_j^T \bx + w_j, \ \ \mbox{for} \ j=1,2,\dots,m, 
\end{equation}
where $\ba_j$ is the $j$-th measurement vector and $w_j$ describes the additive error associated with the $j$-th measurement, which may be due to modeling error or stochastic noise.  Initial breakthrough results in CS established that sparse vectors $\bx$ having no more than $k < n$ nonzero elements can be exactly recovered (in noise-free settings) or reliably estimated (in noisy settings) from a collection of only $m=O(k\log n)$ measurements of the form \eqref{eqn:obs} using, for example, ensembles of randomly generated measurement vectors whose entries are iid realizations of certain zero-mean random variables (e.g., Gaussian) -- see, for example, \cite{Candes:07:Dantzig} as well as numerous CS-related efforts at {\tt dsp.rice.edu/cs}.

While many of the initial efforts in CS focused on purely randomized measurement vector designs and considered recovery of arbitrary sparse vectors, several powerful extensions to the original CS paradigm have been investigated in the literature.  One such extension allows for additional flexibility in the measurement process, so that information gleaned from previous observations may be employed in the design of future measurement vectors.  Formally, such \emph{adaptive sensing} strategies are those for which the $j$-th measurement vector $\ba_j$ is obtained as a (deterministic or randomized) function of previous measurement vectors and observations $\{\ba_{\ell},y_{\ell}\}_{\ell=1}^{j-1}$, for each $j=2,3,\dots,m$.  Non-adaptive sensing strategies, by contrast, are those for which each measurement vector is independent of all past (and future) observations. The randomized measurement vectors typically employed in CS settings comprise an example of a non-adaptive sensing strategy.  Adaptive sensing techniques have been shown beneficial in sparse inference tasks, enabling an improved resilience to measurement noise relative to techniques based on non-adaptive measurements (see, for example, \cite{Ji:08:BCS, Castro:08:Needles, Bashan:08, Haupt:11:DS, Haupt:09:CDS, Newstadt:10, Bashan:11, Iwen:12, Malloy:12, Arias-Castro:11, Haupt:12:CDS, Wei:12, Singh:12, Singh:13, Castro:12} as well as the summary article \cite{Haupt:11:Chapter} and the references therein) and further reductions in the number of compressive measurements required for recovering sparse vectors in noise-free settings \cite{Indyk:11, Price:12}.

Another powerful extension to the canonical CS framework corresponds to the exploitation of additional \emph{structure} that may be present in the locations of the nonzeros of $\bx$.  To formalize this notion, we first define the support ${\cal S} = {\cal S}(\bx)$ of a vector $\bx = [x_1 \ x_2 \ \dots \ x_n]^T$ as
\begin{equation}
{\cal S}(\bx) \triangleq \{i: x_i \neq 0\},
\end{equation}
and note that, in general, the support of a $k$-sparse $n$-dimensional vector corresponds to one of the $n \choose k$ distinct subsets of $\{1,2,\dots,n\}$ of cardinality $k$. The term \emph{structured sparsity} describes a restricted class of sparse signals whose supports may occur only on a (known) subset of these $n\choose k$ distinct subsets.  Generally speaking, knowledge of the particular structure present in the object being inferred can be incorporated into sparse inference procedures, and for certain types of structure this can result either in a  reduction in the number of measurements required for accurate inference, or improved estimation error guarantees, or both (see, e.g., \cite{Zhang:09:Structure, Baraniuk:10:Model, Stojnic:09}, as well as the recent survey article \cite{Duarte:11:Structure} on structured sparsity in compressive sensing).

To the best of our knowledge, our own previous work \cite{Soni:11:Tree} was the first to identify and quantify the benefits of using adaptive sensing strategies that are tailored to certain types of structured sparsity, in noisy sparse inference tasks.  Specifically, the work \cite{Soni:11:Tree} established that a simple adaptive compressive sensing strategy for \emph{tree-sparse} vectors could successfully identify the support of much weaker signals than what could be recovered using non-adaptive or adaptive sensing strategies that were agnostic to the structure present in the signal being acquired.  Subsequent efforts by other authors have similarly identified benefits of adaptive sensing techniques tailored to other forms of structured sparsity in noisy sparse inference tasks \cite{Singh:12, Singh:13, Rao:13:Tree}.  

The primary aim of this effort is to establish the optimality of the strategy analyzed in \cite{Soni:11:Tree}, by identifying the fundamental performance limits associated with the task of support recovery of tree-sparse signals from noisy measurements that may be obtained adaptively.  For completeness, and in an effort to put these results into a broader context, we also identify here the performance limits associated with the same support recovery task in settings where measurements are obtained non-adaptively using randomized (Gaussian) measurement vector ensembles, as in the initial efforts in CS.  We begin by formalizing the notion of tree-structured sparsity, and reviewing the results of \cite{Soni:11:Tree}. 

\subsection{Adaptive Sensing of Tree Sparse Signals}

\begin{figure}[t]
\centering
\includegraphics[width=0.95\linewidth]{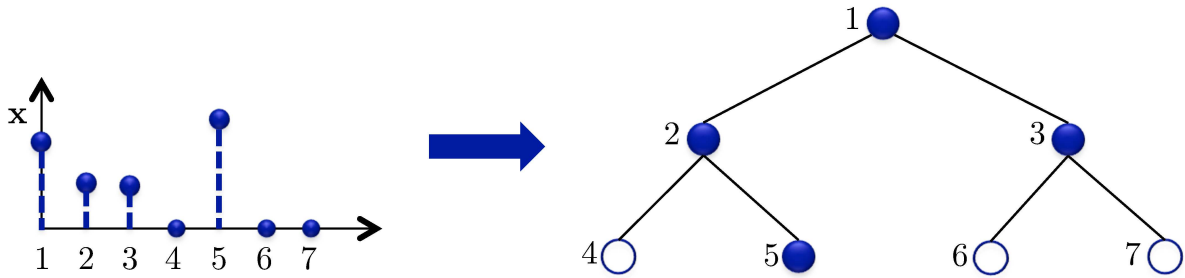} 
\caption{A signal $\bx\in\mathbb{R}^7$ (left) that is $4$-tree sparse in an underlying binary tree having $7$ nodes (right). The support ${\cal S}(\bx) = \{1,2,3,5\}$ corresponds to a rooted connected subtree of the underlying tree.}
\label{fig:tree}
\end{figure}

Tree sparsity essentially describes the phenomenon where the nonzero elements of the signal being inferred exhibit clustering along paths in some known underlying tree.  For the purposes of our investigation here, we formalize the notion of tree sparsity as follows. Suppose that the set $\{1,2,\dots,n\}$ that indexes the elements of $\bx\in\mathbb{R}^n$ is put into a one-to-one correspondence with the nodes of a known tree of degree $d\geq 1$ having $n$ nodes, which we refer to as the \emph{underlying tree}. We say that a vector $\bx$ is \emph{$k$-tree sparse} (with respect to the underlying tree) when the indices of the support set ${\cal S}(\bx)$ correspond, collectively, to a rooted connected subtree of the underlying tree.  
In the sequel we restrict our attention to $n$-dimensional signals that are tree sparse in a known underlying \emph{binary} tree ($d=2$), though our approach and main results can be extended, in a relatively straightforward manner, to underlying trees having degree $d>2$.   For illustration, Figure~\ref{fig:tree} depicts a graphical representation of a signal that is $4$-tree sparse in an underlying complete tree of degree $2$ with $7$ nodes.

Tree sparsity arises naturally in the wavelet coefficients of many signals including, in particular, natural images (see, for example, \cite{Mallat:08:Wavelet, Crouse:98, Romberg:01}), and this fact has motivated several investigations into CS inference techniques that exploit or leverage underlying tree structure in the signals being acquired \cite{Duarte:05:Fast, La:05:Tree, Zhang:09:Structure, Baraniuk:10:Model, Som:12}.  More aligned with our focus here are several prior efforts that have examined specialized \emph{sensing} techniques, designed to exploit the inherent tree-based structure present in the wavelet-domain representations of certain signals in various application domains.  The work \cite{Panych:94:Wavelet}, for example, examined dynamic MRI applications where non-Fourier (in this case, wavelet domain) encoding is employed along one of the spatial dimensions, and proposed a sequential sensing strategy that acquires observations of the wavelet coefficients of the object being observed in a ``coarse-to-fine'' (i.e., top-down, in the wavelet representation) manner. The work \cite{Seeger:08:CSBED} compared a coarse-to-fine direct wavelet coefficient sensing approach to a sensing approach based on Bayesian experimental design in the context of an imaging application.  More recently, \cite{Deutsch:09} proposed a top-down adaptive wavelet sensing strategy in the context of compressive imaging and provided an analysis of the sample complexity of such strategies in noise-free settings, but did not investigate how such procedures would perform in noisy scenarios; see also \cite{Averbuch:12}.  Motivated by these existing efforts, the essential aim of the authors' own prior work \cite{Soni:11:Tree} was to assess the performance of such strategies in noisy settings; for completeness, we summarize the approach and main results of that work here.

Let us assume, for simplicity, that the signal $\bx$ being acquired is tree sparse in the canonical (identity) basis, though extensions to signals that are tree sparse in any other orthonormal basis (e.g., a wavelet basis) are straightforward.  Noisy observations of $\bx$ are obtained according to \eqref{eqn:obs} by projecting $\bx$ onto a sequence of adaptively designed measurement vectors, each of which corresponds to a basis vector of the canonical basis, and we assume that each measurement vector has unit norm.   Now, to simplify the description of the procedure, we introduce some slightly different notation to index the individual observations.  Specifically, rather than indexing observations by the order in which they were obtained as in \eqref{eqn:obs}, we instead index each measurement according to the index of the basis vector onto which $\bx$ is projected, or equivalently here, according to the location of $\bx$ that was observed.  To that end, let us denote by $y_{(j)}$ the measurement obtained by projecting $\bx$ onto the vector $\mathbf{e}_j$ having a single nonzero in the $j$-th location for any $j\in\{1,2,\dots,n\}$.  

Now, begin by specifying a threshold $\tau\geq 0$, and by initializing a support estimate $\widehat{\cal S}=\emptyset$ and a data structure ${\cal Q}$ (which could be a stack, queue, or simply a set) to contain the index corresponding to the root of the underlying tree.  While the data structure ${\cal Q}$ is nonempty, remove an element $\ell$ from ${\cal Q}$, collect a noisy measurement $y_{(\ell)}$ by projecting $\bx$ onto $\mathbf{e}_{\ell}$, and perform the following hypothesis test.  If $|y_{(\ell)}|\geq \tau$, add the indices corresponding to the children of node $\ell$ in the underlying tree to the data structure ${\cal Q}$ and update the support estimate to include the index $\ell$; on the other hand, if $|y_{(\ell)}| < \tau$, then keep ${\cal Q}$ and $\widehat{\cal S}$ unchanged.  Continue in this fashion, at each step obtaining a new measurement and performing a corresponding hypothesis test to determine whether the amplitude of the coefficient measured in that step was significant.  When the overall procedure terminates it outputs its final support estimate $\widehat{\cal S}$, which essentially corresponds to the set of locations of $\bx$ for which the corresponding measurements exceeded $\tau$ in amplitude.   

\begin{algorithm}[t]
\caption{Adaptive sensing procedure for acquiring signals assumed tree-sparse in a (known) underlying tree.}
\label{alg:tree}
\begin{algorithmic}
\REQUIRE Threshold $\tau \geq 0$; Support Estimate ${\cal S}=\emptyset$,
\STATE \hspace{3.3em} Data Structure ${\cal Q}$ containing the index of the root  of the underlying tree
\WHILE{${\cal Q} \neq \emptyset$}
\STATE Remove an index $\ell$ from ${\cal Q}$
\STATE Collect noisy observation $y_{(\ell)} = \mathbf{e}_{\ell}^T\bx + \mathcal{N}(0,\sigma^2)$ 
\IF{$|y_{(\ell)}| \geq \tau$}
\STATE Add indices corresponding to children of $\ell$ in the underlying tree to ${\cal Q}$ 
\STATE Update support estimate: $\widehat{\cal S} \leftarrow \widehat{\cal S} \cup \ell$
\ENDIF
\ENDWHILE
\ENSURE  Final Support Estimate $\widehat{\cal S}$
\end{algorithmic}
\end{algorithm}
The main result of \cite{Soni:11:Tree} quantifies the performance of this type of sensing strategy for acquiring tree-sparse signals in settings where each measurement is corrupted by additive white Gaussian noise; the overall approach in this context is depicted as Algorithm~\ref{alg:tree}. We provide a restatement of the main result of \cite{Soni:11:Tree}\footnote{We note that we have not attempted to optimize constants in our derivation of Lemma~\ref{lem:prev}, opting instead for simple expressions that better illustrate the scaling behavior with respect to the problem parameters.} here as a Lemma, and provide a proof in the appendix, for completeness. It is worth noting that the choice of data structure ${\cal Q}$ in the procedure implicitly determines the order in which measurements are obtained; our analysis, however, is applicable regardless of which particular data structure ${\cal Q}$ is used. 

\begin{lemmai}\label{lem:prev}
Specify a sparsity parameter $k'\in\mathbb{N}$, intended to be an upper-bound for the true sparsity level of the signal being acquired, and choose any $\delta\in(0,1)$. Set the threshold $\tau$ in Algorithm~\ref{alg:tree} to be
\begin{equation}
\tau = \sqrt{2 \sigma^2 \log \left(\frac{4k'}{\delta}\right)}.
\end{equation}
Now, if the signal $\bx\in\mathbb{R}^n$ being acquired by the procedure is $k$-tree sparse for some $k\geq2$, the specified sparsity parameter $k'$ satisfies $k'\leq \beta k$ for some $\beta\geq 1$, and the nonzero components of $\bx$ satisfy
\begin{equation}\label{eqn:lemthresh}
| x_i | \geq \sqrt{8 \left[ 1+\log\left(\frac{4\beta}{\delta}\right) \right]} \cdot \sqrt{\sigma^2 \log k}, 
\end{equation}
for every $i\in{\cal S}(\bx)$, then with probability at least $1-\delta$ the following are true: the algorithm terminates after collecting $m\leq 2k + 1$ measurements, and the support estimate $\widehat{S}$ produced by the procedure satisfies $\widehat{\cal S} = {\cal S}(\bx)$.
\end{lemmai}

In words, this result ensures that when the magnitudes of the nonzero signal components are sufficiently large -- satisfying the condition specified in \eqref{eqn:lemthresh} -- the procedure depicted in Algorithm~\ref{alg:tree} will correctly identify the support of the tree sparse vector (with high probability), and will do so using no more than $2k + 1$ measurements.

Now, as a simple extension, suppose that we seek to identify the support of a $k$-tree sparse vector, and are equipped with a budget of $m$ measurements, where $m\geq r(2k+1)$ for some integer constant $r\geq 1$.  In this setting, the procedure described above may be easily modified to obtain a total of $r$ measurements (each with its own independent additive noise) at each step. If these replicated measurements are averaged prior to performing the hypothesis test at each step, the results of Lemma~\ref{lem:prev} can be extended directly to this setting.  We formalize this extension here as a corollary.

\begin{cori}\label{cor:prev}
Let $\bx$ be as in Lemma~\ref{lem:prev}, and consider acquiring $\bx$ using a variant of the adaptive tree sensing procedure described in Algorithm~\ref{alg:tree}, where $r\geq 1$ measurements are obtained in each step and averaged to reduce the effective measurement noise prior to each hypothesis test.  Choose $\delta\in(0,1)$ and sparsity parameter $k'\in\mathbb{N}$, and set the threshold $\tau$ as
\begin{equation}\label{eqn:threshr} 
\tau = \sqrt{2 \left(\frac{\sigma^2}{r}\right) \log \left(\frac{4k'}{\delta}\right)}.
\end{equation}
If $\bx$ is $k$-tree sparse for some $k\geq2$, the sparsity parameter $k'\leq \beta k$ for some $\beta\geq 1$, and the amplitudes of the nonzero components of $\bx$ satisfy
\begin{equation}
| x_i | \geq \sqrt{8 \left[ 1+\log\left(\frac{4\beta}{\delta}\right) \right]} \cdot \sqrt{\left(\frac{\sigma^2}{r}\right) \log k}, 
\end{equation}
for every $i\in{\cal S}(\bx)$ then with probability at least $1-\delta$ the following are true: the algorithm terminates after collecting $m\leq r(2k + 1)$ measurements, and the support estimate $\widehat{S}$ produced by the procedure satisfies $\widehat{\cal S} = {\cal S}(\bx)$.
\end{cori}
Note that since $m \leq r(2k+1)$ we have that $1/r \leq 3k/m$ provided $k\geq 1$.  It follows from the corollary that when the sparsity parameter $k'$ does not overestimate the true sparsity level by more than a constant factor (i.e., $\beta\geq 1$ is a \emph{constant}), then a sufficient condition to ensure that the support estimate produced by the repeated-measurements variant of the tree sensing procedure is correct with probability at least $1-\delta$, is that the nonzero components of $\bx$ satisfy
\begin{equation}\label{eqn:suffcond}
| x_i | \geq \sqrt{24 \left[ 1+\log\left(\frac{4\beta}{\delta}\right) \right]} \cdot \sqrt{\sigma^2 \left(\frac{k}{m}\right)\log k},  
\end{equation}
for all $i\in{\cal S}(\bx)$.  Identifying whether any other procedure can accurately recover the support of tree-sparse signals having fundamentally weaker amplitudes is the motivation for our present effort.

\subsection{Problem Statement}\label{ssec:prob}

As stated above, the essential aim of this work is to establish whether the adaptive sensing procedure for tree-sparse signals analyzed by the authors in the previous work \cite{Soni:11:Tree}, and summarized above as Algorithm~\ref{alg:tree} is optimal.  Our specific focus here is on establishing  fundamental performance limits for the support recovery task -- that of identifying the locations of the nonzeros of $\bx$ -- in settings where $\bx$ is $k$ tree-sparse, and when observations may be designed either non-adaptively (e.g., measurement vectors whose elements are random and iid, as in traditional CS) or adaptively based on previous observations. We formalize this problem here.

\subsubsection{Signal Model}

Let ${\cal T}_{n,k}$ denote the set of all unique supports for $n$-dimensional vectors that are $k$-tree sparse in the same underlying binary tree with $n$ nodes.  For technical reasons, we further assume that the underlying trees are \emph{nearly complete}, meaning that all levels of the underlying tree are full with the possible exception of the last (i.e., the bottom) level, and all nodes in any partially full level are as far to the left as possible.  

Our specific focus will be on classes of $k$-tree sparse signals, $2\leq k \leq (n+1)/2$, where each $k$-sparse signal $\bx$ has support ${\cal S}(\bx) \in {\cal T}_{n,k}$, and for which the amplitudes of all nonzero signal components are greater or equal to some non-negative quantity $\mu$.  Formally, for a given underlying tree, fixed sparsity level $k$, and ${\cal T}_{n,k}$ as described above, we define the signal class
\begin{eqnarray}\label{eqn:sigmod} 
{\cal X}_{\mu;{\cal T}_{n,k}} \triangleq \left\{\bx\in\mathbb{R}^n: x_i = \alpha_i \mathbf{1}_{\{i\in T\}}, \ |\alpha_i| \geq \mu > 0, \ T\in {\cal T}_{n,k} \right\},
\end{eqnarray}
where $\mathbf{1}_{\{{\cal B}\}}$ denotes the indicator function of the event ${\cal B}$.  In the sequel, we choose to simplify the exposition by denoting the signal class ${\cal X}_{\mu;{\cal T}_{n,k}}$ using the shorthand notation $ {\cal X}_{\mu,k}$, effectively leaving the problem dimension and specification of the underlying tree (and corresponding set of allowable $k$-tree sparse supports) to be implicit.   As we will see, the conditions required for accurate support recovery of $k$-tree sparse signals as defined above are directly related to the signal amplitude parameter $\mu$.

\subsubsection{Sensing Strategies}
 
We examine the support recovery task under both adaptive and non-adaptive sensing strategies.  The non-adaptive sensing strategies that we examine here are motivated by initial efforts in CS, which prescribe collecting observations using ensembles of randomly generated measurement vectors.  Here, when considering performance limits of non-adaptive sensing, we consider observations obtained according to the model \eqref{eqn:obs}, where each $\ba_j$, $j=1,2,\dots,m$, is an independent random vector, whose elements are iid $\mathcal{N}(0,1/n)$ random variables.  This normalization ensures that each measurement vector has norm one in expectation; that is, $\mathbb{E}\left[\|\ba_j\|_2^2\right] = 1$ for all $j=1,2,\dots,m$.  Our investigation of adaptive sensing strategies focuses on observations obtained according to \eqref{eqn:obs}, using measurement vectors satisfying $\|\ba_j\|_2^2 = 1$, for $j=1,2,\dots,m$, and for which $\ba_j$ is allowed to explicitly depend on $\{\ba_{\ell},y_{\ell}\}_{\ell=1}^{j-1}$ for $j=2,3,\dots,m$, as described above.  

Overall, as noted in \cite{Castro:12}, we can essentially view any (non-adaptive, or adaptive) sensing strategy in terms of a collection $M$ of \emph{conditional distributions} of measurement vectors $\ba_j$ given $\{\ba_{\ell},y_{\ell}\}_{\ell=1}^{j-1}$ for $j=2,3,\dots,m$.   We adopt this interpretation here, denoting by $M_{m,{\rm na}}$ the specific sensing strategy based on non-adaptive Gaussian random measurements described above, and by ${\cal M}_{m}$ be the collection of all adaptive (or non-adaptive) sensing strategies based on $m$ measurements, where each measurement vector is exactly norm one (with probability one).

\subsubsection{Observation Noise}

In each case, we model the noises associated with the linear measurements as a sequence of independent $\mathcal{N}(0,\sigma^2)$ random variables.  We further assume that each noise $w_j$ is independent of the present and all past measurement vectors $\{\ba_{\ell}\}_{\ell=1}^{j}$.  For the non-adaptive sensing strategies we examine here noises will also be independent of future measurement vectors, though by design, future measurement vectors generally \emph{will not} be independent of present noises when adaptive sensing strategies are employed.     

\subsubsection{The Support Estimation Task}

We define a support estimator $\psi$ to be a (measurable) function from the space of measurement vectors and associated observations to the power set of $\{1,2,\dots,n\}$.  In other words, an estimator $\psi$ takes as its input a collection of measurement vectors and associated observations, $\{\ba_j,y_j\}_{j=1}^m$, denoted by $\{\bA_m,\by_m\}$ in the sequel (for shorthand), and outputs a subset of the index set $\{1,2,\dots,n\}$.  We note that any estimator can, in general, have knowledge of the sensing strategy that was employed during the measurement process, and we make that dependence explicit here.  Overall, we denote a support estimate based on observations $\bA_m,\by_m$ obtained using sensing strategy $M$ by $\psi(\bA_m,\by_m; M)$.

Now, under the $0/1$ loss function $d(S_1, S_2) \triangleq \mathbf{1}_{\{S_1\neq S_2\}}$ defined on elements $S_1, S_2 \subseteq \{1,2,\dots,n\}$, the (maximum) risk of an estimator $\psi$ based on sensing strategy $M$ over the set ${\cal X}_{\mu,k}$ is given by
\begin{eqnarray}
\nonumber {\cal R}_{{\cal X}_{\mu,k}}(\psi,M) &\triangleq& \sup_{\bx\in{\cal X}_{\mu,k}} \mathbb{E}_{\bx}\left[d( \psi(\bA_m,\by_m; M), {\cal S}(\bx)) \right]\\
&=& \nonumber \sup_{\bx\in{\cal X}_{\mu,k}} \mathbb{E}_{\bx}\left[\mathbf{1}_{\{\psi(\bA_m,\by_m; M) \neq {\cal S}(\bx)\}}\right]\\
&=& \sup_{\bx\in{\cal X}_{\mu,k}}\mbox{Pr}_{\bx}\left(\psi(\bA_m,\by_m; M)\neq{\cal S}(\bx)\right),
\end{eqnarray}
where $\mathbb{E}_{\bx}$ and $\mbox{Pr}_{\bx}$ denote, respectively, expectation and probability with respect to the joint distribution $\mathbb{P}(\bA_m,\by_m; \bx) \triangleq \mathbb{P}_{\bx}(\bA_m,\by_m)$ of the quantities $\{\bA_m,\by_m\}$ that is induced when $\bx$ is the true signal being observed. In words, the (maximum) risk essentially quantifies the worst-case performance of a specified estimator $\psi$ when estimating the ``most difficult'' element $\bx\in{\cal X}_{\mu,k}$ (here, the element whose support is most difficult to accurately estimate) from observations obtained via sensing strategy $M$.

Now, we define the \emph{minimax risk} ${\cal R}^*_{{\cal X}_{\mu,k},{\cal M}}$ associated with the class of distributions $\{\mathbb{P}_{\bx}: \bx\in{\cal X}_{\mu,k}\}$ induced by elements $\bx\in{\cal X}_{\mu,k}$ and the class ${\cal M}$ of allowable sensing strategies as the infimum of the (maximum) risk over all estimators $\psi$ and sensing strategies $M\in{\cal M}$; that is,
\begin{eqnarray}\label{eqn:mmx}
\nonumber {\cal R}^*_{{\cal X}_{\mu,k},{\cal M}} &\triangleq& \inf_{\psi; M\in{\cal M}} R_{{\cal X}_{\mu,k}}(\psi, M) \\
&=&  \inf_{\psi; M\in{\cal M}} \sup_{\bx\in{\cal X}_{\mu, k}} \mbox{Pr}_{\bx}\left(\psi(\bA_m,\by_m;M) \neq {\cal S}(\bx)\right). 
\end{eqnarray}
In words, the minimax risk quantifies the error incurred by the best possible estimator when estimating the support of the ``most difficult'' element $\bx \in {\cal X}_{\mu,k}$ using observations obtained via any sensing strategy $M\in{\cal M}$. 

\sloppypar Note that when the minimax risk is bounded away from zero, so that ${\cal R}_{{\cal X}_{\mu,k},{\cal M}}^* \geq \gamma$ for some $\gamma > 0$, it follows that regardless of the particular estimator $\psi$ and sensing strategy $M\in{\cal M}$ employed, there will always be at least one signal $\bx \in {\cal X}_{\mu,k}$ for which $\mbox{Pr}_{\bx}\left(\psi(\bA_m,\by_m;M) \neq {\cal S}(\bx)\right) \geq \gamma$.  Clearly, such settings may be undesirable in practice, since in this case we can make no \emph{uniform} guarantees regarding accurate support recovery of signals $\bx\in{\cal X}_{\mu,k}$ -- there will always be some worst-case scenario for which the support recovery error probability will exceed $\gamma$.   Our aim here is to identify these problematic scenarios; formally, we aim to identify signal classes ${\cal X}_{\mu,k}$ of the form \eqref{eqn:sigmod}, parameterized by their corresponding signal amplitude parameters $\mu$, for which the minimax risk will necessarily be bounded away from zero.  

\subsection{Summary of Our Contributions}\label{ssec:summary}

Our first main result analyzes the support recovery task for tree-sparse signals in a non-adaptive sensing scenario motivated by the randomized sensing strategies typically employed in compressive sensing.  We state the result here as a theorem, and provide a proof in the next section.

\begin{thmi}\label{thm:na}
Let ${\cal X}_{\mu,k}$ be the class of $k$-tree sparse $n$-dimensional signals defined in \eqref{eqn:sigmod} where $2\leq k \leq (n+1)/2$, and consider acquiring $m$ measurements of $\bx\in{\cal X}_{\mu,k}$ using the non-adaptive (random, Gaussian) sensing strategy $M_{m,{\rm na}}$. If
\begin{equation}\label{eqn:nanec}
\mu \leq \sqrt{\frac{1-2\gamma}{25}} \cdot \sqrt{\sigma^2\left(\frac{n}{m}\right)\log(k)},
\end{equation}
for some $\gamma\in(0,1/3)$ then the minimax risk ${\cal R}_{{\cal X}_{\mu,k},M_{m,{\rm na}}}^*$ defined in \eqref{eqn:mmx} obeys the bound
\begin{equation}
{\cal R}_{{\cal X}_{\mu,k},M_{m,{\rm na}}}^* \geq \gamma.
\end{equation}
\end{thmi}
As alluded above, the direct implication of Theorem~\ref{thm:na} is that no uniform recovery guarantees can be made for any estimation procedure for recovering the support of tree-sparse signals  $\bx\in{\cal X}_{{\mu},k}$ when the signal amplitude parameter $\mu$ is ``too small.''

Our second main result concerns support recovery in scenarios where adaptive sensing strategies may be employed.  We state this result as Theorem~\ref{thm:ad}, and provide a proof in the next section.
\begin{thmi}\label{thm:ad}
Let ${\cal X}_{\mu,k}$ be the class of $k$-tree sparse $n$-dimensional signals defined in \eqref{eqn:sigmod} where $2\leq k \leq (n+1)/2$, and consider acquiring $m$ measurements of $\bx\in{\cal X}_{\mu,k}$ using any sensing strategy $M\in{\cal M}_{m}$. If
\begin{equation}\label{eqn:adnec}
\mu \leq (1-2\gamma) \sqrt{\sigma^2 \left(\frac{k}{m}\right)},
\end{equation}
for some $\gamma\in(0,1/3)$ then the minimax risk ${\cal R}_{{\cal X}_{\mu,k},{\cal M}_m}^*$ defined in \eqref{eqn:mmx} obeys the bound
\begin{equation}
{\cal R}_{{\cal X}_{\mu,k},{\cal M}_m}^* \geq \gamma. 
\end{equation}
\end{thmi}
Similar to the discussion following the statement of Theorem~\ref{thm:na} above, here we have that that no uniform guarantees can be made regarding accurate support recovery of signals $\bx\in{\cal X}_{\mu,k}$ for small $\mu$.

Table~\ref{table:summary} depicts a summary of our main results in a broader context.  Overall, we compare four distinct scenarios corresponding to a taxonomy of adaptive and non-adaptive sensing strategies for recovering $k$-sparse signals under assumptions of unstructured sparsity and tree sparsity.  For each, we identify (up to an unstated constant) a critical value of the signal amplitude parameter, say $\mu^*$, such that for the support recovery task the minimax risk over the class ${\cal X}_{\mu,k}$ will necessarily be bounded away from zero when $\mu \leq \mu^*$. The conditions for support estimation of \emph{unstructured} sparse vectors listed in Table~\ref{table:summary} are a restatement of some known results, and are provided here (with references) for comparison\footnote{Necessary conditions on the signal amplitude parameter required for exact support recovery from non-adaptive compressive samples (and for unstructured sparse signals) were provided in \cite{Aeron:10}; related efforts along these lines include \cite{Wainwright:09:Lasso, Wainwright:09, Genovese:09, Fletcher:09, Wang:10:Sparse}.  Necessary conditions for exact support recovery using adaptive sensing strategies were provided in \cite{Malloy:11:Sequential} for the case where the number of measurements exceeds the signal dimension ($m>n$), while to the best of our knowledge results of this flavor have not yet been established for the compressive regime (where $m<n$).  Finally, we note that several related efforts have established necessary conditions for weaker metrics of \emph{approximate} support recovery using non-adaptive sensing \cite{Reeves:12, Reeves:13} and adaptive sensing strategies \cite{Arias-Castro:11, Castro:12}. }.  
Our main contributions here are depicted in the bottom row of the table, which correspond to the values identified in equations \eqref{eqn:nanec} and \eqref{eqn:adnec}, respectively (with the leading multiplicative factors suppressed).

\begin{table}[t]
    \caption{Summary of necessary conditions for exact support recovery using non-adaptive or adaptive sensing strategies that obtain $m$ measurements of $k$-sparse $n$-dimensional signals that are either unstructured or tree sparse in an underlying nearly complete binary tree.  For each setting, we state the critical value of $\mu$ such that whenever $\mu$ is smaller than a constant times the stated quantity, the minimax risk over the class of signals ${\cal X}_{\mu,k}$ of the form \eqref{eqn:sigmod} will be strictly bounded away from zero.}
    \centering
    \begin{tabular}{ | c | c | c | }
    \hline
     \diagbox[width = 14em, height = 5em]{Sparsity Model}{Sampling Strategy} & Non-adaptive Sensing & Adaptive Sensing \\ \hline
     & & \\
     & $\sqrt{\sigma^2 \left(\frac{n}{m}\right) \log n}$  & $\sqrt{\sigma^2 \left(\frac{n}{m}\right) \log k}$ \\ 
    Unstructured Sparsity & & \\
    & \cite{Aeron:10}; see also \cite{Wainwright:09:Lasso, Wainwright:09, Genovese:09, Fletcher:09, Wang:10:Sparse} & \cite{Malloy:11:Sequential} (when $m>n$)\\ 
    & & \\ \hline
    & & \\
    & $\sqrt{\sigma^2 \left(\frac{n}{m}\right) \log k}$ & $\sqrt{\sigma^2 \left(\frac{k}{m}\right)}$ \\ 
     Tree Sparsity & & \\
     & Theorem~\ref{thm:na} & Theorem~\ref{thm:ad} \\
    & & \\ \hline
    \end{tabular}\label{table:summary}
\end{table}

Two salient points are worth noting when comparing the necessary conditions summarized in Table~\ref{table:summary} with the sufficient condition \eqref{eqn:suffcond} for the repeated-measurement variant of the adaptive tree sensing procedure of Algorithm~\ref{alg:tree}.  First, the results of Theorem~\ref{thm:ad}, summarized in the lower-right corner of Table~\ref{table:summary}, address our overall question -- the simple adaptive tree sensing procedure described above is indeed nearly optimal for estimating the support of $k$-tree sparse vectors, in the following sense: Corollary~\ref{cor:prev} describes a technique that accurately recovers (with probability at least $1-\delta$, where $\delta$ can be made arbitrarily small) the support of any $k$-tree sparse signal from $m\leq r(2k+1)$ measurements, provided the amplitudes of the nonzero signal components all exceed $c_{\delta} \cdot \sqrt{\sigma^2\left(k/m\right)\log k}$ for some constant $c_{\delta}$.  On the other hand, for any estimation strategy based on any adaptive or non-adaptive sensing method, support recovery will fail (with probability at least $\gamma$) to accurately recover the support of some signal or signals in a class comprised of $k$-tree sparse vectors whose nonzero components exceed $c_\gamma \cdot  \sqrt{\sigma^2\left(k/m\right)}$ in amplitude, for a constant $c_{\gamma}$.  

The second noteworthy point here concerns the \emph{relative} performances of the four strategies summarized in Table~\ref{table:summary}.  Overall, we see that techniques that \emph{either} employ adaptive sensing strategies \emph{or} exploit tree structure in the signal being inferred (but not both) may indeed outperform non-adaptive sensing techniques that do not exploit structure, in the sense that either may succeed in recovering signals whose nonzero components are weaker.  That said, the potential improvement arises only in the logarithmic factor present in the amplitudes, implying that either of these improvements by themselves can recover signals whose amplitudes are weaker by a factor that is (at best) a constant multiple of $\sqrt{\log k/\log n}$.  
On the other hand, techniques that leverage \emph{both} adaptivity \emph{and} structure, such as the adaptive tree sensing strategy analyzed above, can provably recover signals whose nonzero component amplitudes are \emph{significantly} weaker than those that can be recovered via any of the other strategies depicted in the table.  Specifically, in this case the relative difference in amplitudes is on the order of a constant times $\sqrt{k/(n \log n)}$, which could be much more significant, especially in high-dimensional settings.  
The experimental evaluation in Section~\ref{sec:eval} provides some additional empirical evidence along these lines.

\subsection{Relations to Existing Works}

As alluded above, several recent efforts have proposed (e.g., \cite{Duarte:05:Fast, La:05:Tree, Som:12}) and analyzed (e.g., \cite{Zhang:09:Structure, Baraniuk:10:Model}) specialized techniques for estimating tree-sparse signals from non-adaptive compressive samples, each of which are designed to exploit the fundamental connectivity structure present in the underlying signal during the inference task.  The work \cite{Seeger:08:CSBED} was among the first to propose and experimentally evaluate a direct wavelet sensing approach for acquiring and estimating wavelet sparse signals (there, images) in the context of compressive sensing tasks, and the sample complexity of a similar procedure in noise free settings was analyzed in \cite{Deutsch:09},\cite{Averbuch:12}.  These works served as the motivation for our initial investigation \cite{Soni:11:Tree} into the performance of such approaches in noisy settings.

Since our work \cite{Soni:11:Tree} appeared, several related efforts in the literature have investigated adaptive sensing strategies for structured sparse signals.  The work \cite{Singh:12}, for example, examined the problem of localizing block-structured activations in matrices from noisy measurements, and established fundamental limits for this task using proof techniques based on \cite{Tsybakov:09}.  We adopt a similar approach based on \cite{Tsybakov:09} below in the proof of one of our main results.  A follow-on work \cite{Singh:13} examined a more general setting, that of support recovery of signals whose supports correspond to (unions of) smaller clusters in some underlying graph.  That work assumed that the clusters comprising the signal model were such that they could be organized into a (nearly balanced) hierarchical clustering having relatively few levels.  While this model is quite general, we note that the class of tree sparse signals we consider here comprise a particularly difficult (in fact, nearly pathological!) scenario for the strategy of \cite{Singh:13}; indeed, the tree-sparse case comprises one example of a problematic scenario identified in \cite{Singh:13} where that approach ``does not significantly help when distinguishing clusters that differ only by a few vertices.''  

It is interesting to note that different structure models can give rise to different thresholds for localization from non-adaptive measurements.  We note, for example, that the thresholds identified in \cite{Singh:12} for localizing block-sparse signals using non-adaptive compressive measurements are weaker than the corresponding threshold we identify in Theorem~\ref{thm:na} here for localizing tree-sparse signals\footnote{Specifically, the results of \cite{Singh:12} imply (adapted to the notation we employ here) that accurate localization of block-sparse signals is impossible when the nonzero signal components have amplitudes smaller than a constant times $\sqrt{\sigma^2\left(\frac{n}{m}\right)\max\left\{\frac{1}{k^{1/2}},\frac{\log n}{k}\right\}}$.}.  This difference arises as a direct result of the different signal models, and in particular, how these differences manifest themselves in the reduction strategy inherent in the proofs based on the ideas of \cite{Tsybakov:09}.  For the analysis of block-sparse signals in \cite{Singh:12} the reduction to hypotheses that are difficult to distinguish leads to consideration of block-sparse signals that either differ on about $k^{1/2}$ locations or do not overlap at all, while in contrast, the performance limits in our case are dictated by tree sparse signals that can differ on as few as two locations.  Stated another way, the tree-sparse signal model we consider here contains subsets of signals that are necessarily more difficult to discern than does the block-sparse model analyzed in \cite{Singh:12}, and this gives rise to the higher necessary signal amplitude thresholds required for localization using non-adaptive compressive measurements for the tree-sparse model we examine here, as compared with the block-sparse model examined in \cite{Singh:12}.

We also note a recent related work  which proposed a technique for sensing signals that are ``almost'' tree-sparse in a wavelet representation, in the sense that their supports may correspond to disconnected subtrees in some underlying tree \cite{Rao:13:Tree}.  While the sensing strategy proposed in that work was demonstrated experimentally to be effective for acquiring natural images, only a partial analysis of the procedure was provided.  Specifically, \cite{Rao:13:Tree} analyzed their procedure only for the case where the signal supports \emph{do} correspond to connected subtrees in some underlying tree, which was effectively the case analyzed in \cite{Soni:11:Tree}. Further, the analysis in \cite{Rao:13:Tree} did not explicitly quantify the sufficient conditions on the signal component amplitudes for which the procedure would successfully recovery the signal support, stating instead only that $m=2k+1$ measurements were sufficient to recover the support provided the SNR was ``sufficiently large.''

While our focus here is specifically on the support recovery task, we note that the related prior work \cite{Singh:12} also identified fundamental limits for the task of \emph{detecting} the presence of block-structured activations in matrices using adaptive or non-adaptive measurements, and established that signals whose nonzero components are essentially ``too weak'' cannot be reliably detected by any method.  Analogous fundamental limits for the detection of certain \emph{tree-sparse} signals have also been established in the literature. Specifically, in the context of our effort here, the problem examined in \cite{Arias-Castro:08:Trail} may be viewed in terms of identifying the support of (a subset of) tree sparse signals whose nonzero elements have the same amplitude $\mu$, from a total of $m=n$ noisy measurements, corresponding to one measurement per node of the underlying tree.  Interestingly, that work established that all detection approaches (for simple trees with no branching) are unreliable when $\mu < c \sqrt{\sigma^2 (n/m)} = c \sqrt{\sigma^2 }$ for a specified constant $c>0$.  This threshold differs from the lower bound we establish for the support recovery task by only a logarithmic factor. This slight difference may arise from the fact that our tree-sparse model contains many more allowable supports (and therefore, more signal candidates) than the path-based model examined in \cite{Arias-Castro:08:Trail}, or it may be that, (at least for the ``full-measurement'' scenario where $m=n$) the support recovery task is slightly more difficult than the detection task.  A full characterization of this type of detection problem for general tree-sparse signals, in settings where measurements may be compressive ($m < n$) as well as adaptive or non-adaptive, is beyond of the scope of our effort here, and remains an (as yet) open problem.

Finally, while our focus here was specifically on the adaptive tree-sensing strategy and fundamental recovery limits for tree-sparse signals, we note that previous results have established that the necessary conditions for recovery of unstructured sparse signals in the top row of Table~\ref{table:summary} are essentially \emph{tight}, in the sense that there exist sensing strategies and associated estimation procedures in each case that are capable of accurate support recovery of sparse signals whose nonzero components exceed a constant times the specified quantity -- see, for example, \cite{Wainwright:09:Lasso, Candes:09:Near, Aeron:10, Dossal:12:Sharp}, which consider the identification of necessary conditions for support recovery of (unstructured) sparse signals from non-adaptive measurements, and \cite{Malloy:11:Sequential}, which analyzes an adaptive sensing strategy for recovering (unstructured) sparse vectors in noisy settings.  Support recovery of (group) structured sparse signals was also examined recently in \cite{Jacob:09, Obozinski:11:Support, Obozinski:11:Group}.

\subsection{Organization}

The remainder of this paper is organized as follows.  The proofs of our main results, Theorems~\ref{thm:na} and \ref{thm:ad}, are presented in Section~\ref{sec:main}.  In Section~\ref{sec:eval} we provide an experimental evaluation of the support recovery task for tree sparse signals. Specifically, where we compare the performance of the tree sensing procedure described above with an inference procedure based on non-adaptive (compressive) sensing that is designed to exploit the tree structure, as well as with adaptive and non-adaptive CS techniques that are agnostic to the underlying tree structure. We also provide experimental evidence to validate the scaling behavior predicted in \eqref{eqn:suffcond} for a fixed measurement budget. We discuss some natural extensions of this effort, and provide a few concluding remarks, in Section~\ref{sec:conc}.  Several auxiliary results, as well as a proof of Lemma~\ref{lem:prev}, are relegated to the Appendix.

\section{Proofs of Main Results}\label{sec:main}

Our first main result, Theorem~\ref{thm:na}, concerns the support recovery task for tree-sparse signals in a non-adaptive sensing scenario motivated by the randomized sensing strategies typically employed in compressive sensing.  Our analysis here follows a similar strategy as in a recent related effort \cite{Singh:12}, which is based on the general reduction strategy described by Tsybakov \cite{Tsybakov:09}.  Our second main result, Theorem~\ref{thm:ad}, concerns support recovery for tree-sparse vectors in scenarios where adaptive sensing strategies may be employed. Our proof approach in this scenario is again based on a reduction strategy -- we argue (formally) that the support recovery task in this case is at least as difficult as the task of localizing a single nonzero signal component of a vector of reduced dimension, and leverage a result of the recent work \cite{Arias-Castro:11} which examined support recovery from non-adaptive measurements for general (unstructured) sparse signals. 

Before we proceed, we first introduce some notation that will be used throughout the proofs here. For any $T\in{\cal T}_{n,\ell}$ with $1\leq \ell<n$, corresponding to the support of a rooted connected subtree with $\ell$ nodes (in some underlying nearly complete binary tree with $n$ nodes), we define $N(T)$ to be the set of locations in the underlying tree, such that for any $j\in N(T)$ the augmented set $T\cup j$ corresponds to a tree with $\ell+1 \leq n$ nodes that is itself another rooted connected subtree of the same underlying tree. Formally, for $T\in{\cal T}_{n,\ell}$ we define
\begin{equation}
N(T) \triangleq \left\{j \in \{1,2,\dots,n\} \ : \ \{T\cup j\} \in {\cal T}_{n,\ell+1}\right\}.
\end{equation} 
With this, we are in position to proceed with the proofs of Theorems~\ref{thm:na} and \ref{thm:ad}.

\subsection{Proof of Theorem~\ref{thm:na}}

The result of Theorem~\ref{thm:na} quantifies the limits of support recovery for tree sparse signals using non-adaptive randomized sensing strategies.  Our analysis is based on the general reduction strategy proposed by Tsybakov \cite{Tsybakov:09}, and follows a similar approach as that in a recent, related effort that identified performance limits for estimating block-structured matrices from noisy measurements \cite{Singh:12}.

Recall the problem formulation and notation introduced in the previous section, and note that for any set ${\cal X}_{\mu,k}' \subseteq {\cal X}_{\mu,k}$, any estimator $\psi$, and any measurement strategy $M\in{\cal M}$, we have that
\begin{eqnarray}
\sup_{\bx\in{\cal X}_{\mu,k}} \mbox{Pr}_{\bx}\left(\psi(\bA_m,\by_m;M) \neq {\cal S}(\bx)\right) \geq \sup_{\bx\in{\cal X}_{\mu,k}'} \mbox{Pr}_{\bx}\left(\psi(\bA_m,\by_m;M) \neq {\cal S}(\bx)\right),
\end{eqnarray}
where as described above the notation $\mbox{Pr}_{\bx}(\cdot)$ denotes probability with respect to the joint distribution $\mathbb{P}(\bA_m,\by_m; \bx) \triangleq \mathbb{P}_{\bx}(\bA_m,\by_m)$ of the quantities $\bA_m$ and $\by_m$ that is induced when $\bx$ is the true signal being observed.  This implies, in particular, that 
\begin{equation}
{\cal R}^*_{{\cal X}_{\mu,k},{\cal M}} \geq {\cal R}^*_{{\cal X}_{\mu,k}',{\cal M}}
\end{equation}
and it follows that we can obtain valid lower bounds on ${\cal R}^*_{{\cal X}_{\mu,k},{\cal M}}$ by instead seeking lower bounds on the minimax risk over any restricted signal class ${\cal X}_{\mu,k}'\subseteq {\cal X}_{\mu,k}$.  This is the strategy we employ here.  

For technical reasons we address the cases $k=2$ and $3\leq k \leq (n+1)/2$ separately, but the essential approach is similar in both cases.  Namely, for each $k$ we construct a set ${\cal X}_{\mu,k}'$ of signals whose nonzero components have the \emph{same} amplitude $\mu$, and whose supports are ``close'' in the sense that the symmetric difference between supports of any pair of distinct signals in the class is a set of cardinality two. 
In each case these signal classes are of the form
\begin{eqnarray}\label{eqn:Xtstar}
{\cal X}_{\mu,k}'(T^*) \triangleq  \left\{\bx\in\mathbb{R}^n: x_i = \mu \mathbf{1}_{\{i\in T\}}, \ T = T^* \cup j, \ j \in N(T^*) \right\},
\end{eqnarray}
for some (specific) $T^*\in {\cal T}_{n,k-1}$ and $N(T)$ is as defined above. This allows us to reduce our problem to the consideration of a hypothesis testing problem over a countable (and finite) number of elements $\bx\in{\cal X}_{\mu,k}'$.

\subsubsection{Case 1: $k=2$}

We begin by choosing $T^*\in{\cal T}_{n,1}$ to be an element of $\mathcal{T}_{n,1}$ for which $|N(T^*)|=2$, and for this $T^*$ we form the set ${\cal X}_{\mu,2}$ of the form \eqref{eqn:Xtstar} above\footnote{Note that this is a somewhat degenerate scenario -- here, $T^*$ can be chosen to be the set that contains only the index of the root node of the underlying tree. Further, that $k\leq (n+1)/2$ implies $n\geq 3$ here, and since the underlying tree is assumed nearly complete, it follows that the root node has two descendants in the underlying tree.}. It follows from the definition of $N(T^*)$ that ${\cal X}_{\mu,2}'(T^*)$ is a set of signals whose supports are each an element of ${\cal T}_{n,2}$, and since each nonzero element has amplitude exactly equal to $\mu$, it follows that every $\bx\in{\cal X}_{\mu,2}'(T^*)$ is also an element of the class of signals ${\cal X}_{\mu,2}$ defined in \eqref{eqn:sigmod} when $k=2$.  Thus, we have overall that ${\cal X}_{\mu,2}'(T^*) \subset {\cal X}_{\mu,2}$.  Now, our approach is to obtain lower bounds on the minimax risk ${\cal R}^*_{{\cal X}_{\mu,2},{\cal M}}$ when ${\cal M}=\{M_{m,{\rm na}}\}$ by considering the minimax risk over the set ${\cal X}_{\mu,2}'(T^*)$, which ultimately corresponds to assessing the error performance of a hypothesis testing problem with two simple hypotheses.

Our analysis relies on a result of Tsybakov \cite[Theorem 2.2]{Tsybakov:09}, which provides lower-bounds on the minimax probability of error for a binary hypothesis testing problem.  We state that result here as a lemma.
\begin{lemmai}[Tsybakov]\label{lem:tsy2}
Let $\mathbb{P}_0, \mathbb{P}_{1}$ be probability distributions (on a common measurable space) for which the Kullback-Leibler (KL) divergence of $\mathbb{P}_0$ from $\mathbb{P}_1$ satisfies $K(\mathbb{P}_1,\mathbb{P}_0) \leq \alpha < \infty$. Then, the minimax probability of error over all (measurable) tests $\psi$ that map observations to an element of the set $\{0,1\}$, given by
\begin{equation}
p_{e,1} \triangleq \inf_{\psi} \max_{j=0,1} \mbox{Pr}_j\left(\psi \neq j\right), 
\end{equation}
where $\mbox{Pr}_j\left(\cdot \right)$ denotes probability with respect to the distribution $\mathbb{P}_j$ induced on the observations when hypothesis $j$ is the correct hypothesis, obeys the bound
\begin{equation}\label{eqn:mpe1}
p_{e,1} \geq \max \left\{\frac{1}{4}\exp\left(-\alpha\right), \frac{1-\sqrt{\alpha/2}}{2}\right\} \geq \frac{1-\sqrt{\alpha/2}}{2}.
\end{equation}
\end{lemmai}

In order to apply this result in our setting, we first need to evaluate the KL divergence $K(\mathbb{P}_1,\mathbb{P}_0)$, where $\mathbb{P}_1$ and $\mathbb{P}_0$ are distributions that characterize our testing problem of identifying which of the two unique elements $\bx_0,\bx_1\in{\cal X}_{\mu,2}'(T^*)$, respectively, was observed.  Now, under the assumption here that the elements of each measurement vector are (iid) Gaussian distributed, we have that the KL divergence of $\mathbb{P}_0$ from $\mathbb{P}_1$ can be expressed in terms of the corresponding probability densities $f_{1} = f_{1}(\{\ba_i,y_i\}_{i=1}^m)$ and $f_{0} = f_{0}(\{\ba_i,y_i\}_{i=1}^m)$ as
\begin{equation}\label{eqn:kl}
K(\mathbb{P}_{1}, \mathbb{P}_{0}) = \mathbb{E}_{1} \left[\log\left(\frac{f_1(\bA_m,\by_m)}{f_0(\bA_m,\by_m)}\right)\right],
\end{equation}
which is just the expectation of the log-likelihood ratio with respect to the distribution $\mathbb{P}_1$.

It follows from the assumptions of our measurement model, specifically that the measurement vectors and noises are mutually independent, that each of the densities $f_p$, $p\in\{0,1\}$, can be factored in the form
\begin{equation}
f_p(\bA_m,\by_m) = \prod_{i=1}^m f(\ba_i) \ f_p(y_i|\ba_i)
\end{equation}
where each $f(\ba_i)$ is multivariate Gaussian density and $f_p(y_i|\ba_i)$ is a (signal-dependent) conditional density of the observation $y_i$ given the measurement vector $\ba_i$.  Note that the conditional densities of $y_i$ given $\ba_i$ are also Gaussian distributed because of the additive noise modeling assumptions.  Overall, the log-likelihood ratio in \eqref{eqn:kl} can be simplified as
\begin{eqnarray}
\nonumber \log\left(\frac{f_1(\bA_m,\by_m)}{f_0(\bA_m,\by_m)}\right) &=& \sum_{i=1}^m\log\left(\frac{f_1(y_i|\ba_i)}{f_0(y_i|\ba_i)}\right)\\
 &=&\nonumber \sum_{i=1}^m \frac{\left(y_i - \ba_i^T\bx_0\right)^2 - \left(y_i - \ba_i^T\bx_1\right)^2}{2\sigma^2}\\
&=& \sum_{i=1}^m \frac{\left(\ba_i^T\bx_0\right)^2 - 2 y_i \ba_i^T\bx_0 - \left(\ba_i^T\bx_1\right)^2 + 2 y_i \ba_i^T\bx_1}{2\sigma^2}.
\end{eqnarray}
Now, using the fact that under the distribution $\mathbb{P}_1$ we have that $y_i = \ba_i^T\bx_1 + w_i$ for $i=1,\dots,m$, and that the noise $w_i$ is zero mean and independent of $\ba_i$, we can simplify the expression \eqref{eqn:kl} as
\begin{eqnarray}
\nonumber K(\mathbb{P}_{1}, \mathbb{P}_{0}) = \mathbb{E}_{1} \left[\sum_{i=1}^m \frac{\left(\ba_i^T(\bx_1 - \bx_0)\right)^2}{2\sigma^2}\right].
\end{eqnarray}

Note that by the construction of ${\cal X}_{\mu,2}'(T^*)$, the vector $\bx_1-\bx_0$ has exactly two nonzero elements, each having amplitude $\mu$ (but with different signs).  It follows that $\ba_i^T(\bx_1-\bx_0)\sim\mathcal{N}(0,2\mu^2/n)$ for each $i=1,\dots,m$, and thus the KL divergence can be expressed simply as
\begin{equation}
K(\mathbb{P}_{1}, \mathbb{P}_{0}) = \frac{m\mu^2}{n\sigma^2}.
\end{equation}
Letting $\alpha = \frac{m\mu^2}{n\sigma^2}$, it is easy to see from \eqref{eqn:mpe1} that if $\alpha \leq 2(1-2\gamma)^2$, or equivalently, if
\begin{eqnarray}
\nonumber \mu &\leq& \sqrt{2(1-2\gamma)^2} \cdot \sqrt{\sigma^2 \left(\frac{n}{m}\right)}\\
&=& \sqrt{\frac{2(1-2\gamma)^2}{\log 2}} \cdot \sqrt{\sigma^2 \left(\frac{n}{m}\right)\log 2},
\end{eqnarray}
for any $\gamma\in(0,1/2)$, then $p_{e,1} \geq \gamma$.

\subsubsection{Case 2: $3\leq k \leq (n+1)/2$}

Analogously to the $k=2$ case, we begin by choosing $T^*\in{\cal T}_{n,k-1}$ to be an element of $T_{n,k-1}$ for which $|N(T^*)|=k$ (the existence of such an element $T^*$ is established by Lemma~\ref{lem:neighbor2} in the appendix) and constructing the set ${\cal X}_{\mu,k}'(T^*)$ to be of the form \eqref{eqn:Xtstar}.  As in the previous case, it follows here that ${\cal X}_{\mu,k}'(T^*) \subset {\cal X}_{\mu,k}$, so our approach here ultimately corresponds to assessing the error performance of a multiple hypothesis testing problem with $k$ simple hypotheses.

We again employ a result of Tsybakov \cite[Proposition 2.3]{Tsybakov:09}, which provides lower-bounds on the minimax probability of error for a hypothesis testing problem deciding among some $L+1$ hypotheses.  We state that result here as a lemma.
\begin{lemmai}[Tsybakov]\label{lem:tsy}
Let $\mathbb{P}_0,\dots,\mathbb{P}_{L}$ be probability distributions (on a common measurable space) satisfying
\begin{equation}
\frac{1}{L} \sum_{j=1}^L K(\mathbb{P}_j,\mathbb{P}_0) \leq \alpha
\end{equation}
with $0 < \alpha < \infty$. Then, the minimax probability of error over all (measurable) tests $\psi$ that map observations to an element of the set $\{0,1,\dots,L\}$, given by
\begin{equation}
p_{e,L} \triangleq \inf_{\psi} \max_{0\leq j\leq L} \mbox{Pr}_j\left(\psi \neq j\right), 
\end{equation}
obeys the bound
\begin{equation}\label{eqn:mpe}
p_{e,L} \geq \sup_{0 < \tau < 1}
\left[\frac{\tau L}{1+\tau L}\left(1 + \frac{\alpha + \sqrt{\alpha/2}}{\log \tau}\right)\right].
\end{equation}
\end{lemmai}
As in the previous case we again need to evaluate KL divergences, this time for pairs of distributions $\mathbb{P}_p$ and $\mathbb{P}_q$ induced by signals $\bx_p,\bx_q\in{\cal X}_{\mu,k}'(T^*)$.  The computation of each KL divergence mirrors the derivation in the previous case; overall, it is straightforward to show that
\begin{equation}
\frac{1}{L} \sum_{j=1}^L K(\mathbb{P}_j,\mathbb{P}_0) = \frac{m\mu^2}{n\sigma^2}.
\end{equation}
Now, note that we can lower-bound the supremum term in the minimum probability of error expression \eqref{eqn:mpe} by evaluating the right hand side for any $\tau\in(0,1)$. Since our test is over $k$ hypotheses we let $L=k-1$ here.  Further, since we consider the case $k\geq 3$ here, we have that $L \geq 2$, so we can choose $\tau = 1/\sqrt{L}\in(0,1)$ to obtain that under the conditions of Lemma~\ref{lem:tsy}, 
\begin{eqnarray}
\nonumber p_{e,L} &\geq& \frac{\sqrt{L}}{1+\sqrt{L}}\left(1 + \frac{\alpha + \sqrt{\alpha/2}}{\log (1/\sqrt{L})}\right)\\
\nonumber &\geq& \frac{1}{2}\left(1 - \frac{(2\alpha + \sqrt{2\alpha})}{\log L}\right)\\
&=& \frac{1}{2}\left(1 - \frac{(2\alpha + \sqrt{2\alpha})}{\log (k-1)}\right).
\end{eqnarray}
Now, note that for any $\gamma\in(0,1/3)$, we have $p_{e,L} \geq \gamma$ whenever $2\alpha + \sqrt{2\alpha} \leq (1-2\gamma)\log(k-1)$, 
or equivalently, whenever $\alpha$ satisfies
\begin{equation}
0 \leq \sqrt{\alpha} \leq \frac{\sqrt{2+8(1-2\gamma)\log(k-1)} - \sqrt{2}}{4},
\end{equation}
which follows from the monotonicity of the function $2\alpha + \sqrt{2\alpha}$ and a straightforward application of the quadratic formula.

As in the previous case, we let $\alpha = \frac{m\mu^2}{n\sigma^2}$  and simplify to obtain that $p_{e,L} \geq \gamma$ whenever
\begin{equation}\label{eqn:peLbracket}
\mu \leq \left[ \ \sqrt{1 + \frac{1}{f_{\gamma, k}}} - \sqrt{\frac{1}{f_{\gamma, k}}} \ \right] \sqrt{\frac{1-2\gamma}{2}} \cdot \sqrt{\sigma^2\frac{n}{m}\log(k-1)},
\end{equation}
where $f_{\gamma, k} = 4(1-2\gamma)\log(k-1)$. Now, for the range of $k$ and $\gamma$ values we consider here we have that $f_{\gamma, k} \geq (4/3)\log(2)$, implying (after a straightforward calculation) that the term in square brackets in \eqref{eqn:peLbracket} is always greater than $0.4 = 2/5$. Thus, we see that $p_{e,L}\geq \gamma$ whenever
\begin{equation}
\mu \leq \sqrt{\frac{2(1-2\gamma)}{25}} \cdot \sqrt{\sigma^2\left(\frac{n}{m}\right)\log(k-1)},
\end{equation}
We make one more simplification, using the fact that $\log(k)/2 < \log(k-1)$ when $k\geq 3$, to claim that if
\begin{equation}
\mu \leq \sqrt{\frac{1-2\gamma}{25}} \cdot \sqrt{\sigma^2\left(\frac{n}{m}\right)\log k},
\end{equation}
then $p_{e,L}\geq \gamma$.

\subsubsection{Putting the Results Together}

In order to combine the results from the previous two cases into one concise form, we first note that for $\gamma\in(0,1/3)$,
\begin{equation}
\sqrt{\frac{1-2\gamma}{25}} < \sqrt{\frac{2(1-2\gamma)^2}{\log 2}}.
\end{equation}
With this, we can claim overall that for any $2\leq k < (n+1)/2$, if for some $\gamma\in(0,1/3)$,
\begin{equation}
\mu \leq \sqrt{\frac{1-2\gamma}{25}} \cdot \sqrt{\sigma^2\left(\frac{n}{m}\right)\log k},
\end{equation}
then the minimax risk over the class ${\cal X}_{\mu,k}$ of $k$-tree sparse signals defined in \eqref{eqn:sigmod} satisfies 
${\cal R}^*_{{\cal X}_{\mu},M_{m,{\rm na}}} \geq \gamma$, as claimed.

\subsection{Proof of Theorem~\ref{thm:ad}} 

Our proof approach in this scenario leverages an essential result from recent efforts characterizing the fundamental limits of support recovery for one-sparse $n$-dimensional vectors \cite{Arias-Castro:11}.  In order to put the results of that work into context here, let us define a class of one-sparse $n$-dimensional vectors as
\begin{equation}\label{eqn:onesparse}
{\cal X}^{(1)}_{\mu} \triangleq \left\{\bx\in\mathbb{R}^n: x_i = \alpha_i \mathbf{1}_{\{i\in T\}}, \ |\alpha_i| \geq \mu > 0, \ T\in [n]\right\},
\end{equation}
where $[n] = \{1,2,\dots,n\}$. Note that we use slightly different notation for the signal class to distinguish it from the tree-sparse classes described above.  In particular, signals in the class \eqref{eqn:onesparse} could have their support on any element of $\{1,2,\dots,n\}$, while in contrast, one-sparse signals that are also tree-sparse must be such that their single nonzero occurs at the root of the underlying tree. 

In terms of the definition \eqref{eqn:onesparse} above, the results of \cite{Arias-Castro:11} (see also the discussion following\cite[Theorem 2]{Davenport:12:CBS}) can be summarized as a lemma.
\begin{lemmai}\label{lem:onesparse}
The minimax risk
\begin{equation}
{\cal R}^*_{{\cal X}^{(1)}_{\mu},{\cal M}_{m}} = \inf_{\psi;M\in{\cal M}_{m}} \ \sup_{\bx\in{\cal X}^{(1)}_{\mu}} \mbox{Pr}_{\bx}\left(\psi(\bA_m,\by_m; M) \neq {\cal S}(\bx)\right)
\end{equation}
over all support estimators $\psi$ and sensing strategies $M\in{\cal M}_{m}$ satisfies the bound
\begin{equation}
{\cal R}^*_{{\cal X}^{(1)}_{\mu},{\cal M}_{m}} \geq \frac{1}{2}\left(1-\sqrt\frac{m\mu^2}{n\sigma^2}\right).
\end{equation}
\end{lemmai}
It follows directly from this result that if
\begin{equation}
\mu \leq (1-2\gamma) \sqrt{\sigma^2\left(\frac{n}{m}\right)}
\end{equation}
for some $\gamma\in(0,1/3)$, then ${\cal R}^*_{{\cal X}^{(1)}_{\mu},{\cal M}_{m}}  \geq \gamma$.   We proceed here by showing (formally) that our problem of interest -- recovering the support of a $k$-tree sparse $n$-dimensional vector using any estimator and any adaptive sensing strategy -- is at least as difficult as recovering the support of a one-sparse vector in some $\widetilde{n}<n$ dimensional space using any estimator and any sensing strategy $M\in{\cal M}_{m}$.  Then, we adapt the result of Lemma~\ref{lem:onesparse} to establish Theorem~\ref{thm:ad}.

We will find it useful in the derivation that follows to introduce an alternative, but equivalent, notation to describe the support estimators and signal supports.  Namely, we associate with any support estimator $\psi$ a corresponding $n$-dimensional vector-valued function $\boldsymbol{\varphi} = [\varphi_1 \ \varphi_2 \ \dots \ \varphi_n]^T$, such that each support estimate $\psi(\bA_m,\by_m; M)$ corresponds to a vector whose elements are given by
\begin{equation}
\varphi_i(\bA_m,\by_m; M) = \mathbf{1}_{\{i\in\psi(\bA_m,\by_m; M)\}},
\end{equation}
for $i=1,2,\dots,n$. Similarly, we can interpret the signal support ${\cal S}(\bx)$ of any vector $\bx$ in terms of an $n$-dimensional binary vector  $\mathbf{S}(\bx) = [S_1(\bx) \ S_2(\bx) \ \dots  \ S_n(\bx)]^T$ with elements
\begin{equation}
S_i(\bx) = \mathbf{1}_{\{i\in{\cal S}(\bx)\}},
\end{equation}  
for $i=1,\dots,n$. 

As in the proof of Theorem~\ref{thm:na}, for any fixed $2\leq k \leq (n+1)/2$ we choose $T^*\in{\cal T}_{n,k-1}$ to be an element of ${\cal T}_{n,k-1}$ for which $|N(T^*)|=k$, and let ${\cal X}_{\mu,k}'(T^*)$ be of the form \eqref{eqn:Xtstar}. Now, observe
\begin{eqnarray}
\nonumber \sup_{\bx\in{\cal X}_{\mu,k}} \mbox{Pr}_{\bx}\left(\psi(\bA_m,\by_m; M) \neq {\cal S}(\bx)\right)
& \geq& \nonumber  \sup_{\bx\in{\cal X}'_{\mu,k}(T^*)} \mbox{Pr}_{\bx}\left(\psi(\bA_m,\by_m; M) \neq {\cal S}(\bx)\right)\\
\nonumber &=& \sup_{\bx\in{\cal X}'_{\mu,k}(T^*)} \mbox{Pr}_{\bx}\left(\cup_{i=1}^n \left\{ \varphi_i(\bA_m,\by_m; M) \neq S_i(\bx)\right\}\right)\\
&\geq& \sup_{\bx\in{\cal X}'_{\mu,k}(T^*)} \mbox{Pr}_{\bx}\left(\cup_{i\in{\cal I}} \left\{ \varphi_i(\bA_m,\by_m; M) \neq S_i(\bx)\right\}\right),
\end{eqnarray}
where in the last line ${\cal I}$ is any subset of $\{1,2,\dots,n\}$.  In particular, this implies that
\begin{eqnarray}\label{eqn:compto}
{\cal R}^*_{{\cal X}_{\mu,k},{\cal M}_{m}}  \geq \inf_{\boldsymbol{\varphi}; M\in {\cal M}_{m}} \ \sup_{\bx\in{\cal X}'_{\mu,k}(T^*)} \mbox{Pr}_{\bx}\left(\cup_{i\in N(T^*)} \left\{ \mathcal{E}_{i}\right\}\right),
\end{eqnarray} 
where $\mathcal{E}_{i}$ is the event  $\{\varphi_i(\bA_m,\by_m) \neq S_i(\bx)\}$. Now, since for any signal $\bx\in{\cal X}'_{\mu, k}(T^*)$ the collection $\{S_i(\bx)\}_{i\in N(T^*)}$ contains exactly one `$1$' and $k-1$ zeros, it follows that the right hand side of \eqref{eqn:compto} is equivalent to the minimax risk associated with the task of recovering the support of a one-sparse $|N(T^*)|$-dimensional vector whose single nonzero element has amplitude $\mu$, in settings where measurements can be obtained via any (possibly adaptive) sensing strategy $M\in{\cal M}_{m}$.   Thus, we can employ the result of Lemma~\ref{lem:onesparse} to conclude that 
\begin{equation}
{\cal R}^*_{{\cal X}_{\mu,k},{\cal M}_{m}} \geq \frac{1}{2}\left(1-\sqrt\frac{m\mu^2}{|N(T^*)|\sigma^2}\right).
\end{equation}
Finally, since $|N(T^*)|=k$, it follows that if for any $\gamma\in(0,1/3)$ we have
\begin{equation}
\mu \leq (1-2\gamma) \sqrt{\sigma^2 \left(\frac{k}{m}\right)},
\end{equation}
then ${\cal R}^*_{{\cal X}_{\mu,k},{\cal M}_{m}} \geq \gamma$, as claimed.

\section{Experimental Evaluation}\label{sec:eval}

In this section we provide several experimental evaluations to validate our theoretical results, and to illustrate the performance improvements that can be achieved in the support recovery task using the adaptive tree sensing procedure.   

In our first experiment we investigate the performance of the tree-sensing approach analyzed here, as the underlying signal dimension increases, and compare the performance of the tree sensing approach with three other strategies from the literature.  Overall, we evaluate four sensing and support estimation strategies, each of which corresponds to one of the four scenarios identified in Table~\ref{table:summary} (adaptive vs. non-adaptive sensing, and structured vs. unstructured sparsity).   The support estimation strategies based on non-adaptive sensing that we evaluate here each utilize measurements obtained according to the model \eqref{eqn:obs}, where measurement vectors are independent Gaussian random vectors with iid ${\cal N}(0,1/n)$ elements.  They are
\begin{itemize}
\item (non-adaptive sensing, unstructured sparsity) a Lasso-based strategy that, from $m$ non-adaptive Gaussian random measurements, first obtains an estimate $\widehat{\bx}_{\rm Lasso}$ according to
\begin{equation}
\widehat{\bx}_{{\rm Lasso},\lambda} = \arg \min_{\bx \in \mathbb{R}^n} \frac{1}{2} \|\by_m - \bA_m\bx\|_2^2 + \lambda \|\bx\|_1,
\end{equation}   
for a constant $\lambda>0$, then forms a corresponding support estimate according to $\widehat{\cal S}_{{\rm Lasso},\lambda} = {\cal S}(\widehat{\bx}_{{\rm Lasso},\lambda})$, and 
\item (non-adaptive sensing, tree sparsity) a Group Lasso-based approach that first identifies an estimate $\widehat{\bx}_{\rm GLasso}$ according to
\begin{equation}
\widehat{\bx}_{{\rm GLasso},\lambda} = \arg \min_{\bx \in \mathbb{R}^n} \frac{1}{2} \|\by_m - \bA_m\bx\|_2^2 + \lambda\sum_{G\in{\cal G}} \|\bx_G\|_2,
\end{equation}   
for a constant $\lambda>0$, where $\bx_G$ denotes the sub vector of $\bx$ indexed by elements in the set $G\subseteq\{1,2,\dots,n\}$ and ${\cal G}$ is the (pre-specified) set of hierarchically overlapping groups which enforce tree structure (see, e.g., \cite{Zhao:09:CAP, Jenatton:10:Proximal}), then forms a support estimate according to $\widehat{\cal S}_{{\rm GLasso},\lambda} = {\cal S}(\widehat{\bx}_{{\rm GLasso},\lambda})$.
\end{itemize}
The adaptive sensing strategies we evaluate are
\begin{itemize}
\item (adaptive sensing, unstructured sparsity) the near-optimal adaptive compressive sensing strategy proposed and analyzed in \cite{Malloy:12:NOACS}, and
\item (adaptive sensing, structured sparsity) the repeated-measurement variant of the adaptive tree sensing approach in Algorithm~\ref{alg:tree} above.
\end{itemize}

We consider overall three different scenarios, corresponding to three different values of the problem dimension ($n=2^{8}-1$, $n=2^{10}-1$, and $n=2^{12}-1$, chosen so that the underlying trees in each case are complete), and in each case we evaluate the performance of each approach over a range of signal amplitude parameters $\mu$, as follows.  In each of $100$ trials we first generate a random $n$-dimensional tree-sparse signal with $k=16$ nonzero components of amplitude $\mu$.  We construct the signals here so that all nonzero components are non-negative, for simplicity, and to facilitate direct comparison with the procedure analyzed in \cite{Malloy:12:NOACS}.  We fix $m=4(2k+1)$ and apply each of the procedures described above (with additive noise variance $\sigma^2=1$), and assess whether it correctly identifies the true support by comparing the support estimate obtained by the procedure with the true support of the tree signal.  The final empirical probabilities of support recovery error for each approach (and each fixed $n$ and $\mu$) were obtained by averaging results over the $100$ trials.  

For completeness, we mention a few additional details regarding our implementations here.  First, for the Lasso-based approaches based on Gaussian measurements, a new independent measurement ensemble was generated to obtain measurements in each trial, but the same measurement vectors and corresponding measurements are used for both of the approaches in a given trial.  Further, since each of the Lasso-based approaches relies on specification of a tuning (regularization) parameter $\lambda$, when evaluating the performance of those approaches we swept over the range of allowable $\lambda$ values, obtaining for each a support estimate as above, and we declare the approach a success if the correct support estimate is identified for \emph{any} value of $\lambda$.  We also note that due to implementation and machine precision issues, the estimates $\widehat{\bx}$ obtained by the Lasso-based estimation strategies may not be exactly sparse; in the experiments we obtained support estimates for each of the Lasso-based estimators by identifying the sets of locations where the corresponding reconstructed signal component amplitudes exceeded $\mu/3$.  The Lasso-based procedures were implemented here using the Sparse Modeling Software (SpaMS), available online at {\tt spams-devel.gforge.inria.fr}.

Our choice of $m$ corresponds to $r=4$ in the repeated-measurement variant of the tree-sensing procedure of Algorithm~\ref{alg:tree}. The threshold for this approach was obtained according to \eqref{eqn:threshr} using $\delta=0.01$ and $\beta=1$.  Note that this choice of $\beta$ corresponds to an instance where the true underlying sparsity level is known prior to implementing the procedure; we afford the procedure of \cite{Malloy:12:NOACS} the same prior knowledge of sparsity level.  Further, strictly speaking, the approach in \cite{Malloy:12:NOACS} does not fit the unit-norm measurement model of \eqref{eqn:obs}, but instead imposes a global constraint on the measurement ensemble of the form $\sum_j \|\ba_j\|_2^2 \leq m$.  In this more general interpretation, the parameter $m$ may be viewed not as the number of measurements per se, but instead as a ``sensing energy'' budget.  Nevertheless, we note that in implementation, each measurement prescribed by the method in \cite{Malloy:12:NOACS} could be synthesized either using a \emph{collection} of measurements obtained using measurement vectors that satisfy $\|\ba_j\|_2^2 = \epsilon$ for all $j$ and some (small) $\epsilon>0$, or equivalently, by appropriately adjusting the effective noise variance per measurement.  We used the latter approach here when implementing the method in \cite{Malloy:12:NOACS}, along with one additional modification. Namely, we note that the procedure in \cite{Malloy:12:NOACS} may not satisfy the sensing energy constraint with equality, in effect leaving some sensing energy unused, which may lead to sacrificed performance. Here, we account for this by explicitly rescaling (increasing) the energy allocations at each step so that the overall sensing energy constraint is satisfied with equality. 

\begin{figure*}[t]
\centering
\includegraphics[width=0.31\linewidth]{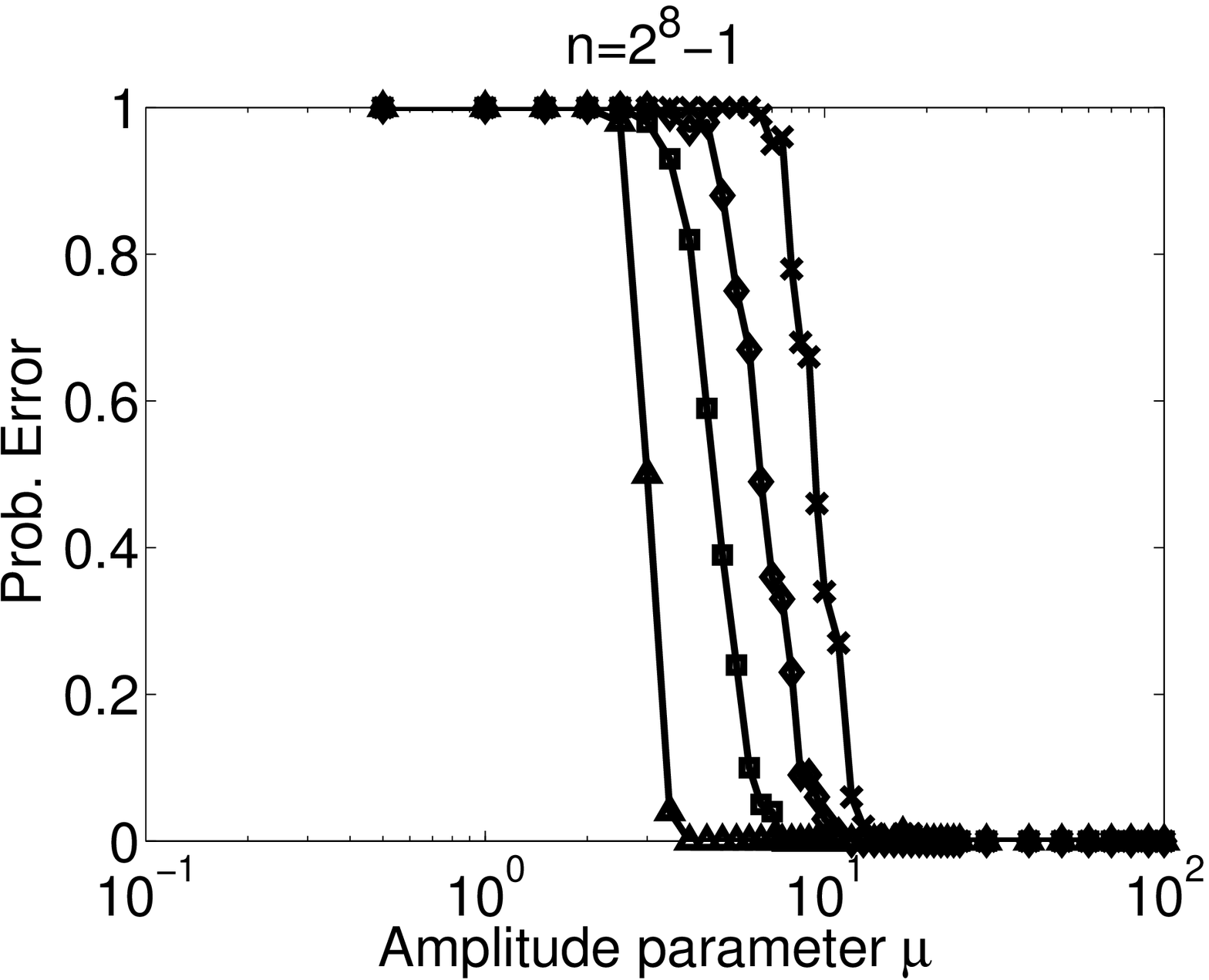}\hspace{1em}   
\includegraphics[width=0.31\linewidth]{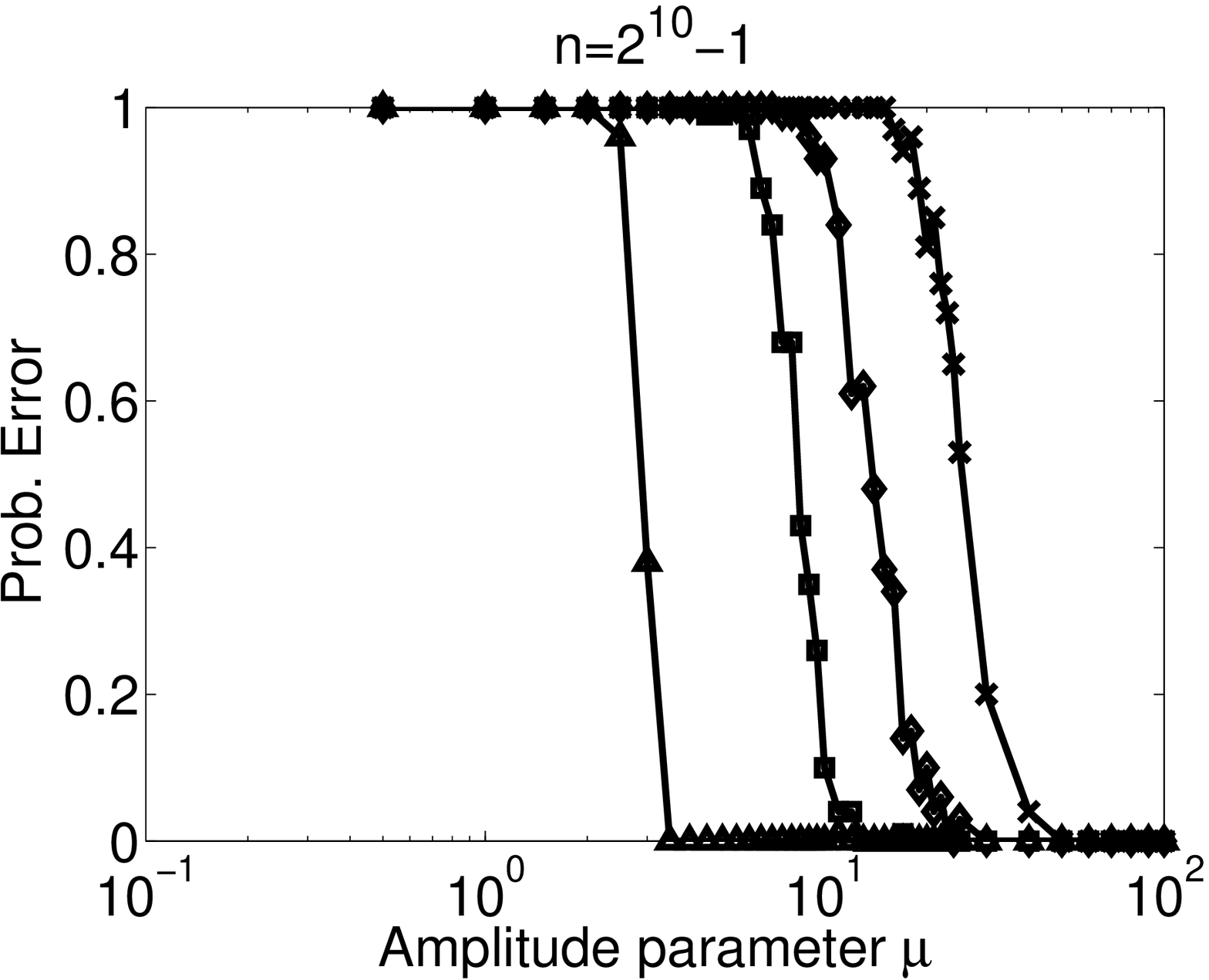} \hspace{1em}
\includegraphics[width=0.31\linewidth]{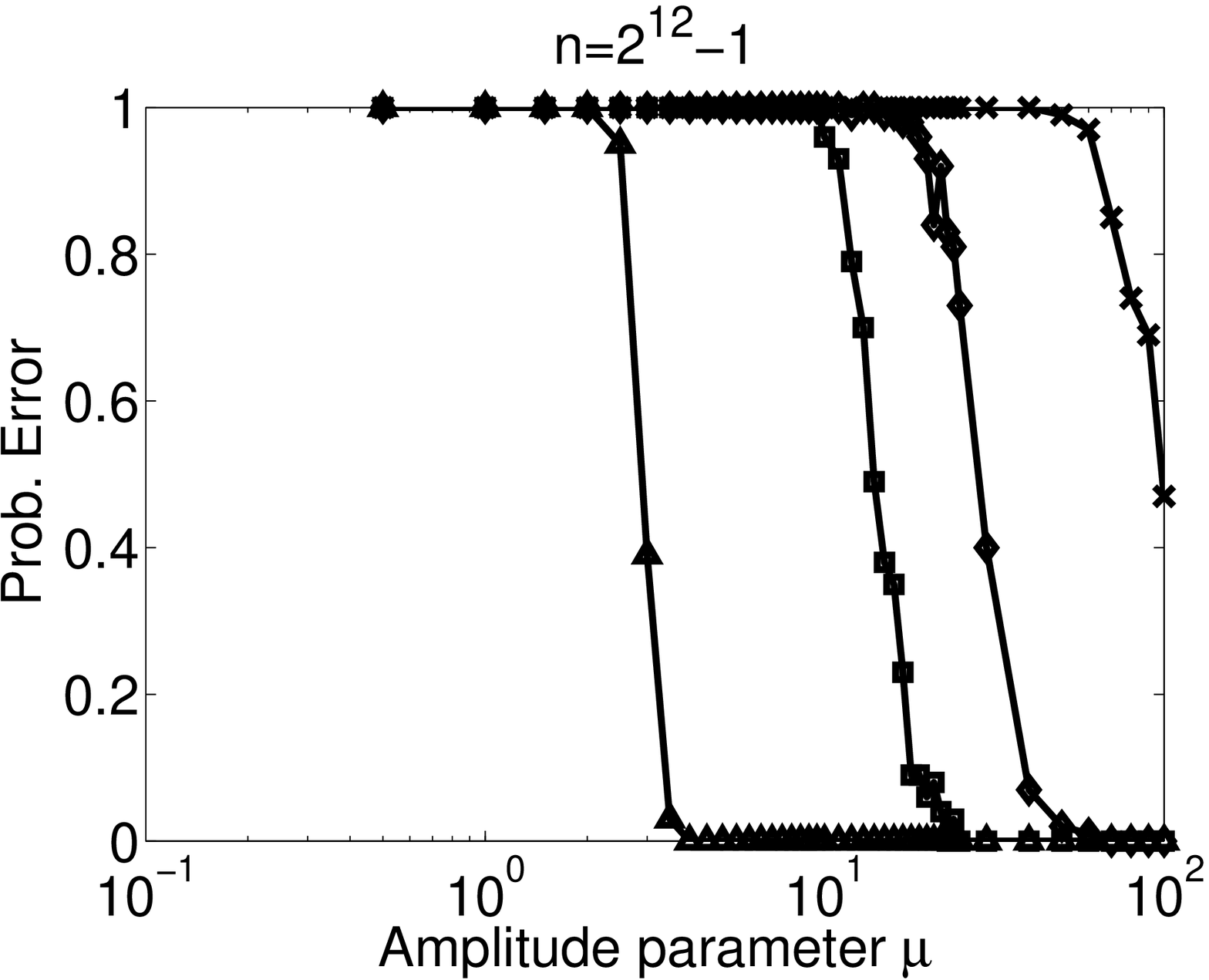}\\
\caption{Empirical probability of support recovery error as a function of signal amplitude parameter $\mu$ in three different problem dimensions $n$. In each case, four different sensing and support recovery approaches -- the adaptive tree sensing procedure described here ($\triangle$ markers); the adaptive compressive sensing approach of \cite{Malloy:12:NOACS} ($\square$ markers); a Group Lasso approach for recovering tree-sparse vectors ($\diamond$ markers), and a Lasso approach for recovering unstructured sparse signals ($\boldsymbol{\times}$ markers) -- were employed to recover the support of a tree-sparse signal with $16$ nonzeros of amplitude $\mu$. The proposed tree-sensing procedure outperforms each of the other methods, and exhibits performance that is unchanged as the problem dimension increases.}
\label{fig:exp}
\end{figure*}

Figure~\ref{fig:exp} depicts the results of this experimental comparison, where for each value of $n$ and for each method we plot the empirical probability of support recovery error, averaged over the $100$ trials, as a function of the (logarithm of the) signal amplitude parameter $\mu$. Here, the tree-sensing procedure corresponds to the curve with triangle ($\triangle$) markers, the adaptive CS approach of \cite{Malloy:12:NOACS} is shown with square ($\square$) markers, the Group Lasso approach is shown with diamond ($\diamond$) markers, and the Lasso approach is shown with the $\boldsymbol{\times}$ markers.  

A few interesting points are worth noting here.  First, as expected, the adaptive tree-sensing procedure outperforms each of the other approaches in each of the three scenarios.  Indeed, the performance of the four approaches follows a fairly intuitive ordering -- the tree-sensing procedure performs best, followed by the adaptive sensing strategy of \cite{Malloy:12:NOACS}, then the Group-Lasso based approach that exploits tree-structure in the inference task (but uses non-adaptive sensing), and finally, the Lasso-based approach that uses non-adaptive sensing, and does not exploit tree structure.   Overall, the results suggest that either utilizing adaptive sensing or exploiting tree structure (alone) can indeed result in techniques that outperform traditional CS, but even more significant improvements are possible when leveraging adaptivity and structure together, confirming our claim in the discussion in Section~\ref{sec:intro}.

Further, it is interesting to note that the performance of the tree-sensing procedure is \emph{unchanged} as the problem dimension increases, in agreement with the result of the result of Corollary~\ref{cor:prev}, where the sufficient condition on $\mu$ that ensures accurate support recovery does not depend on the ambient dimension $n$.  By comparison, the performance of each of the other approaches degrades as the problem dimension increases -- a ``curse of dimensionality'' suffered by each of these other techniques.  While our experimental results only compare problems across $4$ orders of magnitudes, the relative performance differences will become much more significant here as the problem dimension becomes even larger\footnote{We chose these representative problem sizes here, in part, because of computational limitations associated with implementing the Lasso-based experiments on larger problem sizes.  By comparison, our tree-sensing procedure executes in under $1$ second in {\tt MATLAB} on our desktop system, even for problem sizes where $n\sim2^{27}$.}.  

For completeness, we note that the the result of Corollary~\ref{cor:prev}, with the specific parameter choices utilized in our experimental setup, ensures that accurate support recovery (with probability at least $1-\delta=0.99$ here) occurs when $\mu\geq 6.2$.  Here, we observe that accurate support recovery occurs for the tree-sensing procedure for slightly weaker signals whose component amplitudes $\mu$ satisfy $\mu \geq 3.5$.  Of course, the condition identified in Corollary~\ref{cor:prev} is only a \emph{sufficient} condition, and as stated in the discussion in Section~\ref{sec:intro}, we made little effort to optimize the constants associated with the sufficient conditions here, opting instead for results of a simple functional form.  Nevertheless, even with the bounding we employed in our proof, the conditions we identified are fairly representative of the behavior of the procedure in practice.

Our second experimental evaluation is designed to investigate the scaling behavior predicted by the theoretical guarantees we provide in Corollary~\ref{cor:prev} -- namely, that accurate support estimation is achievable provided the nonzero signal components satisfy the condition given in \eqref{eqn:suffcond}.  To that end, we provide a \emph{phase transition} plot for our approach that depicts, for a measurement budget $m_{\rm max}$, whether the tree sensing procedure results in accurate support recovery of $k$ tree-sparse signals whose nonzero amplitudes each have amplitude $\mu$ (for varying parameters $k$ and $\mu$).  More specifically, for this experiment we fix the signal dimension to be $n = 2^{16}-1$ and the noise variance $\sigma^2=1$, we choose $m_{\rm max} = 1000$.  Then, for each choice of the pair $(k,\mu)$ chosen from a discretization of the space $\{1,2,\dots,400\} \times [0,17]$ we implement $100$ trials of the following experiment: generate a random $k$-tree sparse signal having nonzero components with amplitude $\mu$, and implement the tree-sensing strategy described in Corollary~\ref{cor:prev} with threshold $\tau$ as in \eqref{eqn:threshr}, where $r = \lfloor m_{\rm max}/(2k+1)\rfloor$, $k'=k$, and $\delta=0.01$.  We then record, for each choice of sparsity and amplitude parameter, how many of the trials resulted in successful support recovery.

\begin{figure}[t]
\centering
\includegraphics[width=0.6\linewidth]{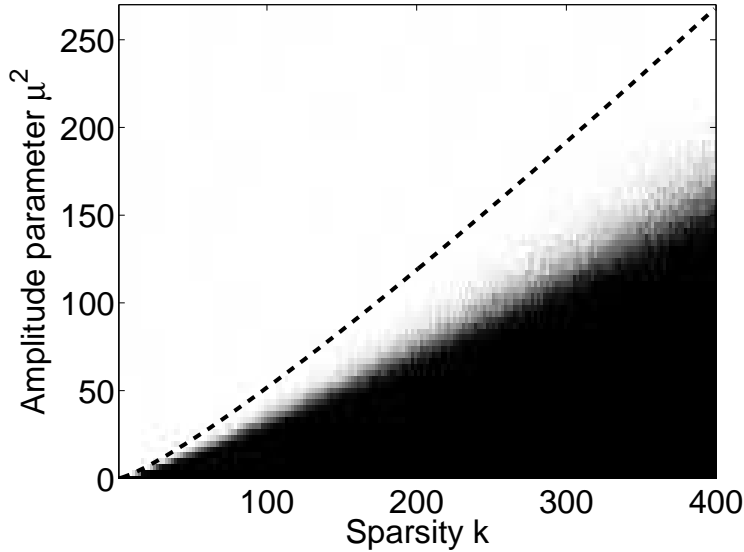}   
\caption{Empirical probability of successful support recovery for the tree-sensing procedure of Algorithm~\ref{alg:tree} as a function of signal sparsity $k$ and squared signal component amplitude $\mu^2$, for a fixed measurement budget.  Here, the light and dark regions correspond to settings where the empirical probability of correct support recovery (averaged over $100$ trials) is nearly one or nearly zero, respectively. The dashed line corresponds to the threshold above which our theoretical results guarantee correct support recovery with probability at least $0.99$.  The empirical results here appear to validate our theoretical predictions for this scenario (see text for specific simulation details).}
\label{fig:exp2}
\end{figure}

The results in Figure~\ref{fig:exp2} depict, for a range of sparsity and amplitude parameters, the fraction of the trials in which the support was exactly identified.  Here, the black regions correspond to the value $0$ and white regions to the value $1$; in words, the dark regions of the plot depict regimes where most or all of the trials failed to successfully identify the true support, the white regions depict regions where the support was accurately identified in a large fraction of the trials, and the grey regions depict the ``transition'' region, where the fraction of trials in which the support was accurately identified is between $0$ and $1$.  

We note that, given our choice of $\delta=0.01$, we expect that the probability of successful support recovery for the tree-sensing procedure should be at least $0.99$ provided the condition \eqref{eqn:suffcond} is satisfied. For comparison, we plot this critical value of signal amplitude corresponding to when the condition \eqref{eqn:suffcond} is satisfied with equality for our parameter choices outlined above (which imply, in particular, that $\beta=1$) in Figure~\ref{fig:exp2} as a dashed line.  From this, we surmise that the $(k,\mu^2)$ pairs depicted by points above the dashed line do indeed correspond to regions where nearly all of the trials resulted in successful support recovery, though as alluded in the discussion of our first experimental evaluation above, the sufficient condition of \eqref{eqn:suffcond} may be a bit conservative. 

The results of Figure~\ref{fig:exp2} allow us to make one additional comparison with the behavior identified by the sufficient condition \eqref{eqn:suffcond}.  Namely, note that for fixed $m, \beta,$ and $\sigma^2$, we expect from \eqref{eqn:suffcond} that the minimum signal amplitude $\mu$ above which exact support recovery is achieved (with high probability) by the tree sensing procedure should increase in proportion to $\sqrt{k \log k}$.  Now, the results of Figure~\ref{fig:exp2} depict success probability as a function of the \emph{square} of the signal amplitude parameter $\mu^2$ and sparsity level $k$, so in this case, we expect that the line in the $\mu^2$ vs $k$ plane above which successful support recovery occurs with probability at least $0.99$ should be functionally proportional to $k \log k$.  This appears to be the case here -- it looks (at least visually, and for this experimental evaluation) like the line demarcating the transition from the black region to the white region does indeed grow super linearly in $k$, providing some additional (visual) validation of the results of Corollary~\ref{cor:prev}.  

\section{Discussion and Conclusions}\label{sec:conc}

In this section we conclude with a few final thoughts including, in particular, some comments on the philosophical difference between the tree-sensing strategy of Algorithm~\ref{alg:tree} and many existing adaptive sensing strategies, as well as a discussion of the implications of our results here for the task of signal estimation. 

\subsection{Adaptive Sensing Strategies for Structured Sparsity}

It is interesting to note that, to date, binary-search-based sensing strategies have been the essential idea behind most of the adaptive sensing procedures that have been proposed and analyzed for sparse recovery tasks in prior efforts including, for example, the aforementioned compressive binary search efforts \cite{Haupt:09:CBS, Iwen:12, Davenport:12:CBS, Singh:12, Singh:13}; the \emph{distilled sensing} strategy of \cite{Haupt:11:DS} (whose analysis provided the first performance guarantees for adaptive sensing strategies in sparse inference tasks) and its compressive sensing variants \cite{Haupt:09:CDS, Haupt:12:CDS}; and the sequential thresholding technique in \cite{Malloy:11:Sequential, Malloy:12}.  The essential functionality of these strategies amounts to ``sequential rejection,'' in the sense that measurements (either compressive, or ``uncompressed'') are initially obtained over all signal locations, and then focused in subsequent steps onto groups or sets of locations of decreasing size, in an attempt to hone in on the true signal components.  

On the other hand, we note that the tree-sensing strategy in Algorithm~\ref{alg:tree} is fundamentally different, in that it is a \emph{constructive} approach.  Indeed, the essential idea behind the procedure of Algorithm~\ref{alg:tree} is to \emph{construct}, in subsequent steps, an subspace of increasing dimension that well-approximates the signal being acquired.  This seemingly subtle difference turns out to be extremely powerful: when using the constructive approach in an adaptive sensing strategy, measurements can be focused locally onto the subspace where the signal exists essentially from the start of the procedure; in other words, no ``global'' measurements need be obtained. In contrast, the binary-search-based strategies necessarily must allocate measurement resources more broadly at the outset, and then only gradually focus onto the signal subspace as it becomes clear via rejection of (enough of) the subspaces or dimensions where the signal is unlikely to reside.  

This fundamental difference has profound implications in the signal recovery task, especially for very high dimensional problems.  Namely, we saw here that the support of $k$-tree sparse signals can be accurately identified provided their nonzero component amplitudes exceed a constant times $\sqrt{\sigma^2(k/m)\log k}$.  This suggests that the problem becomes more ``difficult'' as the sparsity level $k$ increases (as expected) but, notably, the performance is \emph{independent of the ambient signal dimension} $n$.  This was verified in the experimental evaluation in Section~\ref{sec:eval}.  On the other hand sensing and estimation strategies based on compressive binary search ideas and (for example) the general cluster sparse structure investigated in \cite{Singh:13} require component amplitudes that are at least as large as a constant times $\sqrt{\sigma^2 (n/m)\log\log n}$, and further, no sensing and estimation procedure for that type of cluster structure will provide uniform recovery guarantees for signals whose component amplitudes are smaller than a constant times $\sqrt{\sigma^2 (n/m)}$.  In other words, both the necessary and sufficient conditions for recovery of signals exhibiting the form of cluster sparsity studied in \cite{Singh:13} grow with the ambient dimension $n$, implying that the support recovery task for those problems becomes inherently more difficult as the signal dimension $n$ increases, even if the signal sparsity remains fixed. 

That said, it is worth noting a key difference between the tree-sparse signal models we consider here, and the block- and graph-structured models analyzed in \cite{Singh:12, Singh:13}, that gives rise to this distinction.  Namely, in the settings we examine here we enjoy the benefit of a priori ``partial localization'' information, in the sense that we know at the outset one index (corresponding to the root node of the underlying tree) at which the unknown signal has a non-zero component.  This knowledge, along with the strong spatial regularity imposed by the tree structure, is what enables us to accurately localize tree-sparse signals whose component strengths are independent of dimension.  More generally, it is quite likely that the approaches in \cite{Singh:12, Singh:13}, if equipped with analogous partial localization information, could also enjoy the dimension-independent localization thresholds we identify here for the tree-sparse signal models.  Overall, we note that the tree-sparse model we consider here comprises but one useful form of structured sparsity for which the necessary and sufficient conditions for recovery do not suffer an inherent ``curse of dimensionality;'' full characterization of other forms of structured sparsity that exhibit this favorable characteristic is a fruitful path for future investigations.

\subsection{Implications for Signal Estimation}\label{ssec:estimation}

Finally, we note that in some sparse inference tasks it may be more beneficial to assess performance in terms of achievable mean-square error (MSE), rather than by probability of accurate support identification, as here.  While our focus here was specifically on the support recovery task, we conclude with a brief discussion of our results in the context of these estimation tasks.

Several recent efforts have quantified fundamental lower bounds on the achievable MSE when estimating unstructured $k$-sparse signals using (adaptive, or non-adaptive) measurements obtained according to the model \eqref{eqn:obs}.  Specifically, \cite{Candes:12:Well} established that when estimating unstructured $k$-sparse signals $\bx\in\mathbb{R}^n$ using any estimator $\widehat{\bx}$ based on any non-adaptive sensing strategies $M\in{\cal M}_{\rm na}$ for which the ensemble of measurement vectors satisfies the norm constraint 
\begin{equation}\label{eqn:normc}
\|\bA_m\|_F^2 \leq m,
\end{equation}
where $\|\cdot\|_F^2$ denotes the squared Frobenius norm, the minimax MSE satisfies the bound
\begin{eqnarray}
 \inf_{\widehat{\bx},M\in{\cal M}_{\rm na}}  \ \ \sup_{\bx: |{\cal S}(\bx)| = k} \mathbb{E}\left[\|\widehat{\bx}(\bA_m,\by_m;M)-\bx\|_2^2\right] 
 \geq c \  \sigma^2 \left(\frac{n}{m}\right) k \log n,
\end{eqnarray}
for a specified constant $c>0$.  This result established that noisy CS estimation strategies, such as the Dantzig selector \cite{Candes:07:Dantzig}
are essentially optimal, in the sense that there exist measurement ensembles satisfying \eqref{eqn:normc} such that for any $k$-sparse signal $\bx\in\mathbb{R}^n$, the Dantzig selector produces from $m=O(k\log n)$ measurements an estimate $\widehat{\bx}_{\rm DS}$ satisfying $\|\widehat{\bx}_{\rm DS} - \bx\|_2^2 = O\left( \sigma^2 \left(\frac{n}{m}\right) k \log n\right)$ with high probability.   
The works \cite{Arias-Castro:11} and \cite{Castro:12} considered adaptive sensing strategies $M\in{\cal M}_{\rm ad}$ satisfying norm constraints analogous to \eqref{eqn:normc} in the context of estimating unstructured sparse signals, and showed that in this case the minimax MSE satisfies
\begin{eqnarray}
 \inf_{\widehat{\bx},M\in{\cal M}_{\rm ad}}  \ \ \sup_{\bx: |{\cal S}(\bx)| = k} \mathbb{E}\left[\|\widehat{\bx}(\bA_m,\by_m;M)-\bx\|_2^2\right] 
\geq c'' \  \sigma^2 \left(\frac{n}{m}\right) k,
\end{eqnarray}
where $c''>0$ is another constant.  Overall, the improvement that can be achieved using adaptivity when estimating unstructured sparse signals amounts to at most a constant times a logarithmic factor.  

On the other hand, a simple consequence of our support recovery result implies that adaptive sensing strategies for structured sparse signals can result in significant improvements in estimation MSE, as well.  Note that any accurate sparse support estimation procedure based on compressive measurements can easily be parlayed into an estimation procedure by first identifying the locations of the nonzero elements, and then in a second step, collecting direct measurements of the nonzero components (this point was also noted in \cite{Arias-Castro:11}).  Applying this idea using the  adaptive tree sensing strategy described above for the support estimation task, we can establish a result of the following form. 

\begin{lemmai}
There exists a two-stage (support recovery, followed by direct measurement) adaptive compressed sensing procedure for $k$-tree sparse signals that produces, from $m=O(k)$ measurements, an estimate $\widehat{\bx}$ satisfying 
\begin{equation}
\|\widehat{\bx}-\bx\|_2^2 =O\left(  \sigma^2 \left(\frac{k}{m}\right) k \right)
\end{equation}
with high probability, provided the nonzero signal component amplitudes exceed a constant times $\sqrt{\sigma^2 \left(\frac{k}{m}\right)\log k}$ in amplitude.
\end{lemmai}
We omit the proof.  Note that if this type of approach were used to acquire (and estimate) tree-sparse signals whose nonzero components have equal amplitudes, it follows that the estimate $\widehat{\bx}$ produced would satisfy 
\begin{equation}
\mathbb{E}\left[\|\widehat{\bx}-\bx\|_2^2\right] = O\left(\sigma^2 \left(\frac{k}{m}\right)k \log k\right).
\end{equation}
Somewhat astonishingly, this bound is (up to constants) within a logarithmic factor of the estimation error that would be produced were an oracle to provide the \emph{exact} locations of the nonzero components \emph{before any measurements were obtained}! This argument suggests that accurate estimation approaches (based on adaptive sensing strategies) can be constructed for recovering a broad class of relatively weak tree-sparse signals (i.e., signals having very small individual component amplitudes), and that these strategies could be capable of producing estimators whose errors are on the order of those incurred by oracle-aided sensing strategies.   We defer a more thorough investigation of the performance limits for this tree-sparse signal estimation task to a later effort.

\section{Acknowledgements}

The authors are grateful to the anonymous reviewers for their detailed and thorough evaluations, and in particular, for pointing out some subtle errors in the initial versions of our proofs of our main results, and for helping to clarify the potential implications of the results of \cite{Candes:12:Well}  for achievable minimax MSE for estimation of signals exhibiting structured sparsity.

\appendix
\section{Auxiliary Material}\label{sec:proof}

We first establish a few useful intermediate results that will be used in the proof of Lemma~\ref{lem:prev} as well as in the proofs of our main theorems.  Recall from the discussion in Section~\ref{sec:main} that for any $T\in{\cal T}_{n,\ell}$, corresponding to a tree with $1\leq \ell < n$ nodes that is a rooted connected subtree of some underlying nearly complete binary tree of $n$ nodes, we defined the set $N(T)$ to be the set of locations at which one additional node can be added to the tree described by $T$ to yield a tree with $\ell+1$ nodes that is itself another rooted connected subtree of the same binary tree.  The following lemma provides a bound on the sizes of the sets $N(T)$.

\begin{lemmai}\label{lem:neighbor}
For any $T\in{\cal T}_{n,k}$ with $k<n$, we have that $|N(T)| \leq k+1$. 
\end{lemmai}

\vspace{1em}\begin{proof}
The proof proceeds by induction on $k\leq n$.  The case $k=1$ is a trivial case where ${\cal T}_{n,1}$ contains only a single element corresponding to the index of the root node of the underlying tree. Now, since the underlying tree is binary we have that the number of locations at which one node can be added is less than or equal to $2$. 

Now, for some $k<n$ we assume that $|N(T)|\leq k+1$ for all $T\in{\cal T}_{n,k}$; we aim to show that $N(T')\leq (k+1) + 1$ for all $T'\in{\cal T}_{n,k+1}$.  To that end, we note that any $T'\in {\cal T}_{n,k+1}$ can be expressed as $T' = T \cup j$ for some $T\in{\cal T}_{n,k}$ and $j\in N(T)$.  This implies that $N(T')$ contains all elements in the set $N(T)\setminus j$, as well as the children of the index $j$, of which there are at most two.  Thus, it follows that for any $T'\in{\cal T}_{n,k+1}$ we have that $|N(T')| \leq (|N(T)| - 1) + 2 \leq (k+1)-1+2 = (k+1) + 1$, which is what we intended to show. 
\end{proof}\vspace{1em}

It is worth noting that results of this flavor are somewhat classical.  For example, \cite{Knuth:72:Art1} establishes a related result in a setting where the size of the underlying tree is essentially unconstrained, implying that each node has exactly two children.  This corresponds to a special case of the above result, where the number of locations at which one node may be added is \emph{exactly} $k+1$.  We opt to provide the above proof for completeness, but also to highlight the difference in the setting where the size of the underlying tree is constrained, which is an essential characteristic of our signal model.

Our second intermediate result identifies settings where the bounds obtained above on the cardinality of the set $N(T)$ hold with equality.  For this, we make explicit use of the assumption that the underlying tree be nearly complete; even in this case, the result holds only for signals that are ``sparse enough.''
\begin{lemmai}\label{lem:neighbor2}
For every integer $2\leq k\leq (n+1)/2$, there exists at least one $T^*\in{\cal T}_{n,k-1}$ for which $|N(T^*)|=k$.  
\end{lemmai}
It follows directly from this lemma that there exists a subset ${\cal T}^* = \left\{ T^* \cup j: j\in N(T^*) \right\} \subseteq{\cal T}_{n,k}$ of allowable supports for $k$-tree sparse vectors, for which $|{\cal T}^*|=k$, and the symmetric difference $T_i\Delta T_j \triangleq (T_i \cup T_j) \setminus (T_i \cap T_j)$ satisfies $|T_i\Delta T_j| = 2$ for any pair $T_i,T_j\in{\cal T}^*$ with $i\neq j$.

\vspace{1em}\begin{proof}
Recall that the underlying tree is assumed to be nearly complete, meaning that all levels of the underlying tree are full with the possible exception of the last level, and that nodes in the last level are as far to the left as possible.  Our proof is constructive -- for each $2\leq k\leq (n+1)/2$ we let $T^*$ be the set of indices that corresponds to the $k-1$ nodes in the underlying tree selected in a top-down, left-to-right manner.  

First, note that when $k-1=2^q-1$ for some integer $q\in\mathbb{N}$ the set of indices in $T^*$ will correspond to a \emph{complete} subtree of the underlying nearly complete tree.  Further, the underlying tree must contain all $2^{q}=k$ nodes in the level immediately below the last level filled by the indices in $T^*$; if not, then the total number of nodes in the tree would satisfy $n < (k-1) + k = 2k-1$, which contradicts the condition that $k \leq (n+1)/2$.  Thus, in this case the $N(T^*)$ can be taken to be the $2^q=k$ descendants of the nodes in the last full level of the subtree described by the indices in $T^*$.

For other values of $k$ that cannot be written as integer powers of $2$, the set $T^*$ will not correspond to a complete subtree, but instead, a nearly complete subtree of the underlying tree.  In this case, note that $d^* = d^*(k) = \lfloor \log_2(k-1) \rfloor$ is the depth of the last (partially-filled) level of the subtree corresponding to the $k-1$ elements in $T^*$. It follows that the total number of indices in all of the filled layers of the subtree defined by elements of $T^*$ is $2^{d^*}-1$, and thus, the number of indices in the last partially-filled level is given by $v^* = (k-1) - (2^{d^*}-1)$.  Now, the set $N(T^*)$ can be constructed to contain all of the $2^{d^*}-v^*$ indices in the last partially filled level, plus $2v^*$ indices in the level immediately below that are the descendants of the indices in the last partially filled level.  For this construction, note that
\begin{eqnarray}
\nonumber |N(T^*)| &=& 2^{d^*}-v^* + 2v^*\\
\nonumber &=& 2^{d^*} + (k-1) - (2^{d^*}-1)\\
&=& k.
\end{eqnarray}
Finally, for completeness, we note that the level immediately below the last partially filled level of the subtree described by elements of $T^*$ must indeed contain at least $2v^*$ indices.  If not, then the total number of indices in the underlying tree would be $n < (k-1) + k = 2k-1$, which again contradicts the requirement that $k\leq (n+1)/2$.
\end{proof}\vspace{1em}

\subsection{Proof of Lemma~\ref{lem:prev}}

The proof of Lemma~\ref{lem:prev} relies on the fact that when acquiring a particular signal $\bx$ having support ${\cal S}(\bx)\in{\cal T}_{n,k}$, the support estimate $\widehat{\cal S}$ produced when the adaptive sensing procedure of Algorithm~\ref{alg:tree} terminates will exactly equal the true support when the event
\begin{equation}\label{eqn:evente}
{\cal E}_{\bx} \triangleq \left\{\bigcap_{i\in{\cal S}(\bx)} |y_{(i)}| \geq \tau \right\} \cap \left\{ \bigcap_{j\in N({\cal S}(\bx))} |y_{(j)}| < \tau\right\}
\end{equation}
occurs\footnote{Note that in our proof  we adopt the alternative notation used in our description of the adaptive sensing procedure, where measurements are indexed according to the location that was observed.}. More specifically, the event ${\cal E}_{\bx}$ corresponds to a (valid) instance of the procedure where measurements of $\bx$ are obtained at all locations $\ell \in {\cal S}(\bx) \cup N({\cal S}(\bx))$ and the hypothesis test associated with each measurement produces the correct result, thus resulting in a correct final support estimate.  

In other words, the above discussion establishes that ${\cal E}_{\bx} \subseteq \{\widehat{\cal S} = {\cal S}(\bx)\}$. (Actually, the events ${\cal E}_{\bx}$ and $\{\widehat{\cal S} = {\cal S}(\bx)\}$ can be shown to be equal, though we don't explicitly need this fact for our proof.)  This implies, in turn, that $\{\widehat{\cal S} = {\cal S}(\bx)\}^c \subseteq {\cal E}^c_{\bx}$; thus,
\begin{eqnarray}\label{eqn:errbnd}
\nonumber  \mbox{Pr}_{\bx} \left(\widehat{\cal S} \neq {\cal S}(\bx)\right)
\nonumber &&\leq \mbox{Pr}_{\bx}\left(\left\{\bigcup_{i\in{\cal S}(\bx)} |y_{(i)}| < \tau \right\} \cup \left\{ \bigcup_{j\in N({\cal S}(\bx))} |y_{(j)}| \geq \tau\right\}
\right)\\
&&\leq \sum_{i\in{\cal S}(\bx)} \mbox{Pr}_{\bx} \left(|y_{(i)}| < \tau \right) + \sum_{j\in N({\cal S}(\bx))} \mbox{Pr}_{\bx} \left( |y_{(j)}| \geq \tau\right). 
\end{eqnarray}
The proof proceeds by identifying simple upper bounds for each term in the sum on the right hand side of \eqref{eqn:errbnd}. To that end, note that for $j\in N({\cal S}(\bx))$ we have that $y_{(j)}\sim\mathcal{N}(0,\sigma^2)$. Thus,
\begin{eqnarray}
\nonumber \mbox{Pr}_{\bx} \left( |y_{(j)}| \geq \tau\right) &=& \mbox{Pr}_{\bx} \left(\{ y_{(j)} \geq \tau\} \cup \{y_{(j)} \leq -\tau\}\right)\\
\nonumber &=& 2 \cdot \mbox{Pr}_{\bx} \left(y_{(j)} \geq \tau\right)\\
&\leq& \exp\left(-\frac{\tau^2}{2\sigma^2}\right),
\end{eqnarray}
where the second line follows by symmetry and the fact that the events are disjoint, and the third line utilizes a standard bound on the tail of the Gaussian distribution.

We now consider obtaining bounds on the probabilities of the events $\{|y_{(i)}| < \tau\}$ for $i\in{\cal S}(\bx)$.  Note that for $i\in{\cal S}(\bx)$ we have $y_{(i)} = \alpha_i + w$ where $w\sim\mathcal{N}(0,\sigma^2)$ represents the additive noise for that observation.  Since we placed no condition on the signs of the nonzero elements of $\bx$ we ultimately have to consider two cases to establish our bound.  Consider, first, the case where the nonzero element at location $i$ satisfies $\alpha_i>0$. We have
\begin{equation}
\{|y_{(i)}| < \tau\} = \{-\tau - \alpha_i < w < \tau-\alpha_i\}\subset \{w < \tau-\alpha_i\},
\end{equation}
implying that $\mbox{Pr}_{\bx}\left( \{|y_{(i)}| < \tau\} \right) \leq \mbox{Pr}_{\bx}\left( w < \tau-\alpha_i \right)$. Now, for $\tau < \mu \leq \alpha_i$ we can again employ a standard bound on the tail of the Gaussian distribution to claim
\begin{equation}
\mbox{Pr}_{\bx}\left( \{|y_{(i)}| < \tau\} \right) \leq \exp\left(-\frac{(\alpha_i - \tau)^2}{2\sigma^2}\right).
\end{equation}
Using a similar approach for the case $\alpha_i < 0$ and the same $\tau$, we obtain (after some straightforward computations) that the overall the bound
\begin{eqnarray}
\nonumber \mbox{Pr}_{\bx}\left( \{|y_{(i)}| < \tau\} \right) &\leq& \exp\left(-\frac{(|\alpha_i| - \tau)^2}{2\sigma^2}\right)\\
&\leq& \exp\left(-\frac{(\mu - \tau)^2}{2\sigma^2}\right)
\end{eqnarray}
holds for any $i\in{\cal S}(\bx)$, where the last step follows from the fact that $|\alpha_i| \geq \mu$ for all $i\in{\cal S}(\bx)$.
Thus
\begin{eqnarray}\label{eqn:errbnd2}
\mbox{Pr}_{\bx} \left(\widehat{\cal S}\neq {\cal S}(\bx)\right)  \leq k \exp\left(-\frac{(\mu - \tau)^2}{2\sigma^2}\right) + (k+1)\exp\left(-\frac{\tau^2}{2\sigma^2}\right).
&&
\end{eqnarray}
Note that the leading factor of $k+1$ in the second term of \eqref{eqn:errbnd2} is a consequence of Lemma~\ref{lem:neighbor}.

The last step of the proof is straightforward, and amounts to showing that for any $\delta\in(0,1)$, when
\begin{equation}
\mu = \sqrt{8 \log\left(4/\delta\right)} \cdot \sqrt{\sigma^2 \log k}
\end{equation}
and
\begin{equation}
\tau = \sqrt{2 \sigma^2 \log \left(4k/\delta\right)},
\end{equation}
each of the two terms in the sum in \eqref{eqn:errbnd2} can be upper bounded by $\delta/2$.  Further, it is easy to verify that for these choices $\tau < \mu$ (as required by our proof) whenever $k>1$.  These steps are straightforward, so we omit the details.

\bibliographystyle{IEEEbib}
\bibliography{treebib}

\begin{IEEEbiographynophoto}{Akshay Soni}(S'08) received his B.Tech. degree in Information and Communication technology from Dhirubhai Ambani Institute of Information and Communication Technology (DA-IICT), Gujarat, India, in 2010, and the Masters of Science degree in Electrical Engineering from University of Minnesota, Minneapolis, USA, in 2011. Since then he has been working towards his Ph.D. degree with the Department of Electrical and Computer Engineering, University of Minnesota. His research interests include adaptive compressive sensing, machine learning and statistical learning theory.

Mr. Soni is the recipient of a departmental travel award at the University of Minnesota in 2012. During the summer of 2012 he worked as an intern with Mitsubishi Electric Research Labs (MERL), Cambridge, Massachusetts, USA.
\end{IEEEbiographynophoto}

\begin{IEEEbiographynophoto}{Jarvis Haupt}(S'05--M'09) received the B.S., M.S. and Ph.D. degrees in electrical engineering from the University of Wisconsin--Madison in 2002, 2003, and 2009, respectively. From 2009-2010 he was a Postdoctoral Research Associate in the Department of Electrical and Computer Engineering at Rice University in Houston, TX. He is currently an Assistant Professor in the Department of Electrical and Computer Engineering at the University of Minnesota. His research interests generally include high-dimensional statistical inference, adaptive sampling techniques, statistical signal processing and learning theory, and applications in the biological sciences, communications, imaging, and networks. 

Dr. Haupt is the recipient of several academic awards, including the Wisconsin Academic Excellence Scholarship, the Ford Motor Company Scholarship, the Consolidated Papers Tuition Scholarship, the Frank D. Cady Mathematics Scholarship, and the Claude and Dora Richardson Distinguished Fellowship.  He served as Co-Chair of the Teaching Improvement Program at the University of Wisconsin-Madison for two semesters, and received Honorable Mention for the Gerald Holdridge Teaching Award for his work as a teaching assistant there.  He has also completed technical internships at Georgia Pacific, Domtar Industries, Cray, and L-3 Communications/Integrated Systems, and was a consulting engineer for GMR Research and Technology.
\end{IEEEbiographynophoto}
\end{document}